\RequirePackage{lineno}
\documentclass[prb,preprint,floatfix]{revtex4}

\usepackage{graphicx, nicefrac}
\usepackage{array}
\usepackage{amsmath,amsfonts,amssymb,braket}
\usepackage[ansinew]{inputenc}
\usepackage{color,soul}
\usepackage{calc}
\usepackage{multirow,booktabs}
\usepackage{xcolor,textcomp}
\usepackage{listings}
\usepackage{lineno}

\newcommand{\Figref}[1]{Fig.~\ref{#1}}
\newcommand{\Figureref}[1]{Figure~\ref{#1}}

\newcommand{\ignore}[1]{}

\begin{document}

\title{Electrically driven cascaded photon-emission in a single molecule}

\author
{Katharina Kaiser,$^{1,2\ast}$ Anna Ros{\l}awska,$^{1,3}$ Michelangelo Romeo,$^{1}$\\ 
Fabrice Scheurer,$^{1}$ Tom{\'a}\v{s} Neuman,$^{4,5\ast}$ Guillaume Schull$^{1\ast}$\\
\normalsize{$^{1}$Universit\'{e} de Strasbourg, IPCMS, CNRS, UMR 7504, Strasbourg, France}\\
\normalsize{$^{2}$4th Physical Institute -- Solids and Nanostructures, Georg-August-Universit\"{a}t G\"{o}ttingen}\\
\normalsize{G\"{o}ttingen, 37077, Germany}\\
\normalsize{$^{3}$Max Planck Institute for Solid State Research, Stuttgart, 70569, Germany}\\
\normalsize{$^{4}$Universit\'{e} Paris-Saclay, Institut des Sciences Mol\'{e}culaires d'Orsay (ISMO)}\\
\normalsize{CNRS, UMR 8214, Orsay, 91405, France}\\
\normalsize{$^{5}$Institute of Physics, Czech Academy of Sciences}\\
\normalsize{Cukrovarnick\'{a} 10, Prague, 16200, Czech Republic}\\
\normalsize{$^\ast$To whom correspondence should be addressed:}\\
\normalsize{E-mail: katharina.kaiser@uni-goettingen.de; neuman@fzu.cz; schull@unistra.fr.}
}

\begin{abstract}

Controlling electrically-stimulated quantum light sources (QLS) is key for developing integrated and low-scale quantum devices. The mechanisms leading to quantum emission are complex, as a large number of electronic states of the system impacts the emission dynamics. Here, we use a scanning tunneling microscope (STM) to excite a model QLS, namely a single molecule. The luminescence spectra reveal two lines, associated to the emission of the neutral and positively charged molecule, both exhibiting single-photon source behavior. In addition, we find a correlation between the charged and neutral molecule's emission, the signature of a photon cascade. By adjusting the charging/discharging rate, we can control these emission statistics. This generic strategy is further established by a rate equation model revealing the complex internal dynamics of the molecular junction. 

\end{abstract}

\maketitle

The generation of non-classical states of light is highly relevant to both fundamental science and research into quantum information technologies. In particular, quantum light sources (QLS) such as defect centers in diamond and Si\cite{Kurtsiefer2000PRL,Rogers2014NatCommun,Bradac2019NatCommun,Higginbottom2022Nature,Prabhu2023NatCommun}, quantum dots\cite{Michler2000Science,Garcia2021Science}, or single molecules\cite{ Brunel1999PRL,Lounis2000Nature,Wang2019NatPhys,Murtaza2023OptExpress,Toninelli2021NatMater} are promising platforms for future quantum communication, cryptography, and computing applications\cite{Waks2002Nature,Beveratos2002PRL,OBrien2007Science,Kimble2008Nature}. Electrical driving of these sources\cite{Yuan2002Science,Salter2010Nature,Nothaft2012NatCommun,Mizuochi2012NatPhoton,Bentham2016ApplPhysLett,Lin2017NatCommun}, as opposed to the more prevalent optical pumping schemes, is interesting for practical purposes such as device integration and has been shown to potentially undercut the lower limit of single-photon purity achievable in photoluminescence\cite{Fischer2017NaturePhys,Lin2017NatCommun,Aharonovich2016NatPhoton}. From a fundamental perspective, electrical driving of QLS is highly intriguing because the formation of excited states is mediated through the creation of charged species and can thus entail transitions into multiple states\cite{Mizuochi2012NatPhoton,Lin2017NatCommun,Jiang2023PRL}. This complexity is both a blessing and a curse, since it generally jeopardizes the central goal of developing platforms that can be precisely controlled, while at the same time providing ways to manipulate photon statistics by electro-generation of excited states. A comprehensive understanding of the underlying generation mechanism on the individual emitter level is thus vital to pave the way toward controllably addressing QLS in an all-electrical excitation scheme.\\
Here we use scanning tunneling microscopy-induced luminescence (STML)\cite{Qiu2003Science,Zhang2016Nature,Imada2016Nature,Doppagne2017PRL} in combination with Hanbury Brown-Twiss (HBT) interferometry \cite{Brown1956Nature,Perronet2006EPL,Merino2015NatCommun,Zhang2017NatCommun,Luo2019PRL,Roslawska2020ACSNano} to study and control the electroluminescence of an individual model QLS, namely a single zinc-phthalocyanine (ZnPc) molecule. By electrically pumping the charged excited state of this molecule, we generate the emission of photons of different colours, originating from the radiative decay of the molecular trion and exciton. We show that, while individually both show single photon emission, their crosscorrelation reveals clear signatures of a cascaded photon emission, where the trion announces the exciton. We construct a comprehensive picture of the transition dynamics between the different many-body states that are relevant in the electro-generation of correlated single photons from ZnPc. Notably, we show that the formation of the molecular trion is promoted through the optically dark triplet state which is generally formed with high likelihood in electroluminescence experiments\cite{Yersin2004Book,Nothaft2012NatCommun}. Building upon these insights, we explore the tunability of the excited state dynamics in ZnPc by manipulating the charging rate from the tip, demonstrating that we can gain a nuanced control over the transition rates and relative average population of the many-body states that participate in the excited state formation. This provides a high degree of control over the properties of the single molecule QLS.

\subsection*{Correlated emission from individual ZnPc molecules}

\begin{figure}
\centering
    \includegraphics{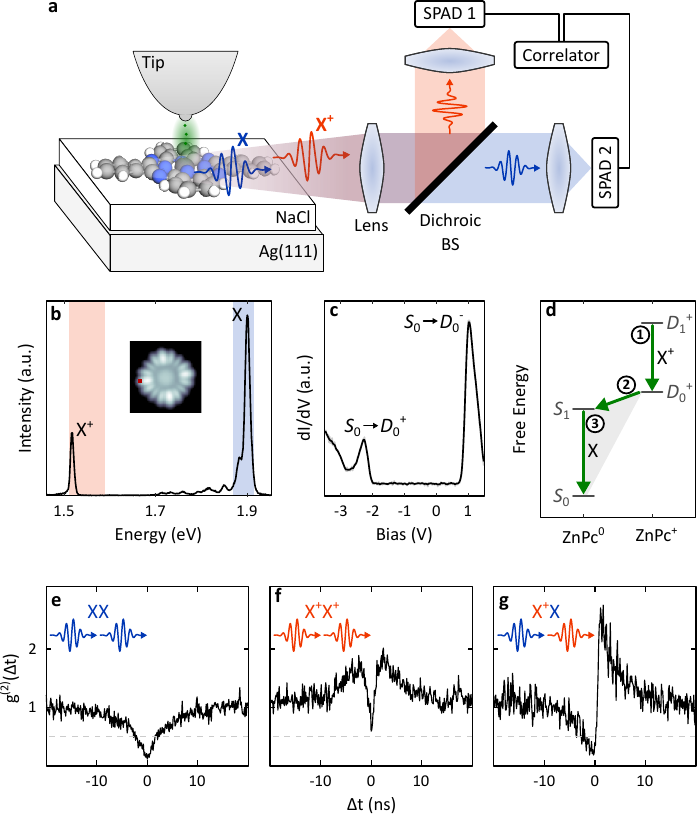}
    \caption{\textbf{Photon-correlation of exciton and trion emission in ZnPc.} (a) Schematic of the setup. The tunnel current between tip and substrate is used to drive exciton and trion electroluminescence from single ZnPc adsorbed on 4ML NaCl/Ag(111). (b) STML spectrum recorded at -3 V, 130 pA (position indicated in the inset, image recorded at $-2.3\,\text{V}$, $5\,\text{pA}$), showing emission from the exciton (X, neutral molecule) and the trion ($\text{X}^+$, charged molecule). The emitted photons are spectrally sorted by a dichroic beam splitter and collected by two single-photon avalanche diodes (SPAD) connected to a correlator in an HBT setup. (c) d$I$/d$V$ on ZnPc/3 ML NaCl/Ag(111), setpoint -3.5 V, 25 pA. (d) Many-body scheme for the exciton formation (grey shaded area) and the cascaded emission process (green arrows). After the molecule has been initiated in $D_1^+$, the $X^+$ transition (1) is followed by a charge-state transition (2) into $S_1$ and a consecutive radiative decay into the molecule's ground state (X, 3). (e--g) Typical $g^{(2)}(\Delta t)$ measurements for the XX and $\text{X}^+\text{X}^+$ (e, f) as well as the $\text{X}^+\text{X}$ correlation (g), recorded at -3 V and 30 pA, binning 128 ps.}
    \label{Fig1}
\end{figure}

\Figureref{Fig1}a shows a sketch of our experimental setup, where the tip of a low-temperature scanning tunneling microscope (STM) is used to excite the luminescence of a single ZnPc molecule decoupled from an Ag(111) substrate by 3--5 monolayers (ML) of NaCl. If not indicated otherwise, the presented data were recorded on 4 ML NaCl. The emitted photons are detected in the far field by a spectrometer and an HBT setup, respectively (a detailed description of the setup as well as the HBT measurements can be found in Supporting Information). The STML spectrum in \Figref{Fig1}b, recorded at a bias voltage of -3 V, shows two distinct peaks at 1.89 eV and 1.52 eV. They can be assigned to the radiative decay of the excited to the ground state of the neutral ($S_1\rightarrow S_0$) and charged ($D_1^+\rightarrow D_0^+$) molecule, respectively, \textit{i.e.}, the radiative decay of the molecular exciton and trion\cite{Doppagne2018Science,Dolezal2021ACSNano,Hung2023PRR}. Excited state formation in STML usually proceeds \textit{via} sequential resonant tunneling events to and from the molecule\cite{Roslawska2018NanoLett,Miwa2019NanoLett,Jiang2023PRL,Hung2023PRR}. In the case of ZnPc/NaCl/Ag(111), the neutral excited state $S_1$ can be efficiently populated at the positive ion resonance (PIR). This relates to the bias voltage ($\approx$ -2.1 V) at which transient positive charging of the molecule from the tip, corresponding to the $S_0\rightarrow D_0^+$ transition, becomes energetically possible, as can be deduced from the d$I$/d$V$ spectroscopy (\Figref{Fig1}c). From $D_0^+$, $S_1$ can be populated by a discharging event from the sample and eventually decays radiatively, leaving the molecule in its neutral ground state ($S_0$) by excitonic emission, a mechanism that has been described in detail elsewhere\cite{Miwa2019NanoLett,Jiang2023PRL,Hung2023PRR}. The population of $D_1^+$, whose radiative decay to $D_0^+$ leads to the trionic emission ($X^+$), follows a similar charging-discharging-scheme that will be discussed in detail later.
\\The intricate interplay of these transitions in the electroluminescence process can be understood by looking at intensity-intensity correlations of the emission, \textit{i.e.} the second order correlation $g^{(2)}(\Delta t)$, defined as
\begin{align*}
    g^{(2)}(\Delta t) &= \frac{\braket{I_{\alpha}(t)I_{\beta}(t+\Delta t)}}{\braket{I_{\alpha}(t)}\braket{I_{\beta}(t)}},
\end{align*}
where $\alpha,\beta\; \in \{\text{X},\text{X}^+$\}, $I_{\alpha,\beta}(t)$ is the intensity recorded at time $t$ for $\alpha$ and $\beta$, respectively, and $\Delta t$ is the time delay between the detection of two photons. \Figref{Fig1}e--g shows correlation measurements of the excitonic (XX) and trionic (X$^+$X$^+$) emission as well as the two emission channels (X$^+$X). The observed anti-bunching in XX, with $g^{2}(0)<0.5$, is a feature of the sub-Poissonian photon statistics of a single photon emitter. In STML, the typical timescale at which anti-bunching is observed has first been associated to the exciton lifetime\cite{Merino2015NatCommun,Zhang2017NatCommun,Dolezal2021ACSNano}. The width of this dip, which here is on the order of a few nanoseconds, in fact non-trivially relates to both the lifetime of the excited state and the time it takes until this excited state is re-populated\cite{Beveratos2001PRA,Roslawska2020ACSNano,Fishman2023PRX}. In STML, these two processes occur at distinctly different timescales. The radiative decay of the optically excited state is expected to occur within a few picoseconds due to the strong electromagnetic field enhancement\cite{Roslawska2022PRX,Dolezal2024NanoLett}, which is beyond the time resolution of the HBT setup (ultimately given by the photodetector's response). The excitation, on the other hand, involves charge transfer through the NaCl layer and is thus limited to nanosecond timescales for molecules adsorbed on 4 ML NaCl\cite{Kaiser2023NatCommun,Steurer2014ApplPhysLett}. The width of the anti-bunching appears to be dominated by the slower process, which in this case is the charge transfer.\\ 
In the X$^+$X$^+$ correlation (\Figref{Fig1}f), the observed anti-bunching is again a fingerprint of single-photon emission and occurs within few nanoseconds. However, unlike in the XX correlation, this is followed by symmetric bunching at around +/-3 ns. This feature is reminiscent of systems in which the photocycle is disrupted by the population of a dark state\cite{Draebenstedt1999PRB,Beveratos2001PRA,Schroeder2021ACSNano}, the nature of which will be discussed later in the manuscript. $g^{(2)}(\Delta t)$ of X$^+$ and X in \Figref{Fig1}g shows clear correlations between the two channels: At negative $\Delta t$, $g^{(2)}(\Delta t)$ shows a dip similar to the XX correlation. In contrast, at positive $\Delta t$ we find a sharp increase in $g^{(2)}(\Delta t)$ close to $\Delta t = 0\,\text{s}$ that converges to $g^{(2)}(\Delta t) = 1$ within approximately 10 ns. This cross-bunching (anti-bunching for negative and bunching for positive $\Delta t$) is a fingerprint of a correlated photon pair in which the radiative decay of the trion announces the radiative decay of the exciton, \textit{i.e.}, an emission cascade\cite{Aspect1980PRL,Moreau2001PRL,He2016NatCommun} within a single molecule. The correlation between the emission from excitons and trions can be rationalized as follows: After initiating ZnPc in its charged excited state $D_1^+$, the radiative decay of the trion (X$^+$) leaves the molecule in its charged ground state $D_0^+$ (\textbf{1} in \Figref{Fig1}d). Upon discharging, the $D_0^+\rightarrow S_1$ transition (\textbf{2}) is energetically available and thus, after the radiative decay of the trion, exciton formation upon electron transfer from the substrate is likely. The subsequent radiative decay (X) of $S_1$ to the neutral ground state $S_0$ (\textbf{3}) occurs within a short time, leading to bunching at $\Delta t > 0\,\text{s}$. After the system has returned to its neutral ground state, X$^+$ emission is suppressed until $D_1^+$ is re-populated, leading to anti-bunching in the X$^+$X correlation at $\Delta t < 0\,\text{s}$. The observed cross-bunching is thus indeed evidence for a trion-exciton cascade generating correlated photons of different colours from a single ZnPc molecule in STML.

\subsection*{Tuning the photon statistics in a cascaded emission process}

\begin{figure}
\centering
    \includegraphics[width = \textwidth]{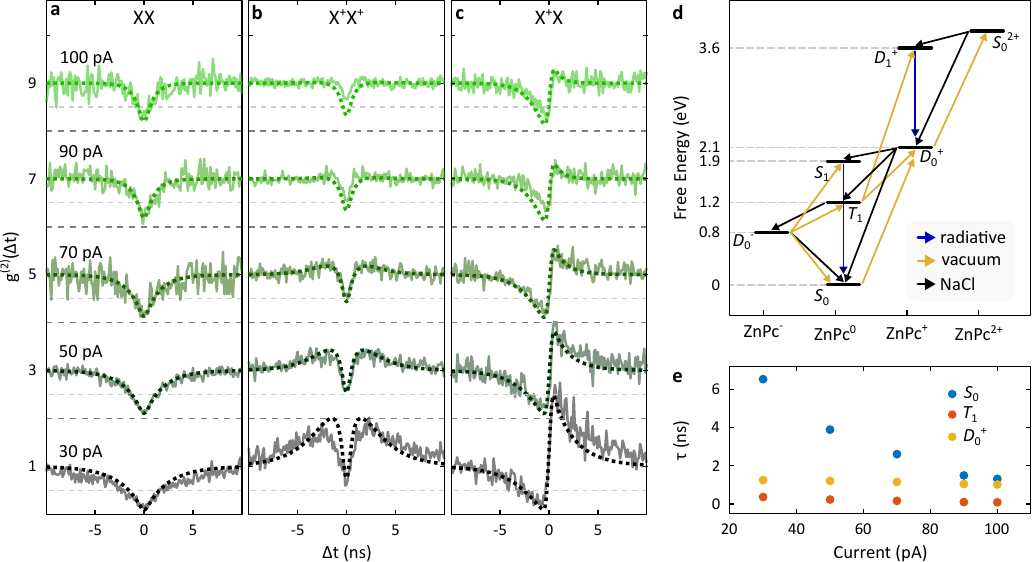}
    \caption{\textbf{Current dependent exciton and trion dynamics.} (a-c) Correlation measurements of the excitonic and trionic emission showing the (a) XX, (b) X$^+$X$^+$ and (c) X$^+$X correlations recorded at -3 V for different current setpoints from 30 pA (black) to 100 pA (light green). Solid curves: experimental data; dashed curves: fits from the theoretical model. The histograms are each offset by 2 for visibility. The horizontal dashed lines indicate the 0 and 0.5 values of the normalized counts, respectively. (d) Free Energy diagram of the many-body states and transitions that govern the charging and discharging dynamics involved in exciton and trion formation. Black arrows: tunneling events through the NaCl, yellow arrows: tunneling events through the vacuum gap, blue arrows: radiative transitions. (e) Lifetimes of $S_0$, $T_1$, and $D_0^+$ as a function of current, as extracted from the model.}
    \label{Fig2}
\end{figure}

As described above, the dynamics of excited state formation in STML appears to be dictated by charge-state transitions from tip and substrate. Thus, changing the charging dynamics, \textit{e.g.} by changing the tip-sample distance and with it the tunnel current, should directly influence the statistics of excitonic and trionic emission. Since the tunneling probability between molecule and metallic substrate (sample-mediated transitions) is solely determined by the NaCl thickness and the energetic alignment of the molecular states that are involved in the transitions within the NaCl band gap \cite{Steurer2014ApplPhysLett,Kaiser2023NatCommun}, it remains practically unaffected by changes in the tip-molecule distance. Charge-state transitions by tunneling through the vacuum barrier between tip and molecule (tip-mediated transitions) are on the other hand, at a given bias voltage, exclusively determined by the tip-molecule distance. Hence, by varying the tip-molecule distance, we deliberately tune certain charge-state transitions while leaving others unaltered. To investigate their influence on the trion-exciton dynamics and the underlying photon statistics, we recorded the XX, X$^+$X$^+$, and X$^+$X correlations as a function of tunnel current (\Figureref{Fig2}a--c). Upon increasing the tunnel current, \textit{i.e.} increasing the probability for tip-mediated transitions, the anti-bunching dip in the XX correlation narrows down (\Figref{Fig2}a). A qualitatively similar trend is also observable in the X$^+$X cross-bunching (\Figref{Fig2}c) at negative $\Delta t$. At the same time, the anti-bunching width in X$^+$X$^+$ (\Figref{Fig2}b) does not visibly change, indicating that this feature is not or only weakly influenced by the tip-mediated rates. The  bunching in both X$^+$X$^+$ and X$^+$X is most pronounced at low current (30 pA) and decreases as the current increases. Note that the general progression of $g^{(2)}(\Delta t)$ is independent of the tip position and the exact applied bias voltage provided that $D_1^+$ can be populated (see Supporting Information S2 for additional data).

To relate the observed features to a given excitation pathway, we derive a theoretical description of the statistics of correlated emission based on the possible many-body transitions. We design a rate-equation model containing the many-body states required to simulate the experiment. Besides considering the $S_0$, $S_1$ and $D_0^+$, $D_1^+$ states directly participating in the photon emission,  we also include in the model possible intermediate states involved in the charging and discharging processes. These include the triplet state $T_1$ of the neutral molecule, the ground state $S_0^{2+}$ of the doubly charged molecule, and the ground state $D_0^-$ of the negatively charged molecule. The dynamics and steady-state values of the populations of these states are then governed by rates indicating the charging and discharging processes and direct decay of the excitons. We use this model to fit the correlation curves at different tunnel currents, taking into account the limited time resolution of the detectors. To that end, we fit the relative tip-molecule distance - reflecting the changes in vacuum tunneling probability - that scales the tip-mediated tunneling rates. This is eventually the only fitted value in each curve. The remaining parameters, including the substrate-mediated rates, the plasmon-enhanced exciton (trion) decay rate, and other constants defining the tip-mediated rates are selected based on values from the literature and the present experiment and are kept fixed for all effective tip-sample distances. The model is summarized in the diagram of \Figref{Fig2}d, where the many-body states are connected with arrows marking the respective rates (processes) considered. More details about the model and the values of its parameters are discussed in Supporting Information Section S3. Using the model, we can now access the transition rates that govern the features in the correlation curves. In particular, this allows us to determine the lifetimes of the mediating states (see Supporting Information Section S3 and S4 for more details). \Figref{Fig2}e shows the lifetimes of $S_0$, $T_1$ and $D_0^+$ as a function of current. Of these three states, $T_1$ exhibits the shortest lifetimes on the order of a few hundred picoseconds, which is related to a fast transition from $T_1$ to $D_0^+$ (see Supporting Information Section S4 for the different transition rates as a function of current and further discussion). Since this corresponds to a tip-mediated transition, the $T_1$ lifetime reduces by about a factor of 4 with increasing current. Within the probed current-range, the transitions that govern the depopulation of $D_0^+$ are sample-mediated transitions to the three neutral ZnPc$^0$ states (\Figref{Fig2}d), such that the $D_0^+$ lifetime remains essentially unaffected by changes in the tip-mediated transition rates. $S_0$, on the other hand, can only be depopulated by tip-mediated transitions, such that, similarly to $T_1$, its lifetime changes by about a factor of 5 from initially 6.5 ns at 30 pA to 1.3 ns at 100 pA.\\ 
The changes in lifetime of the states that mediate exciton and trion formation directly relate to the progression of the observed features in the correlation curves. The anti-bunching width in XX, for instance, is governed by the lifetimes of the mediating states $S_0$ and $D_0^+$. Since of these two only the $S_0$ lifetime changes with current, the narrowing of the dip in \Figref{Fig2}a can be directly associated to a reduction of the $S_0$ lifetime. The features in the X$^+$X$^+$ correlation are in fact more complicated to assign since a direct population of $D_1^+$ through a charge-state transition from $S_0$, as previously discussed\cite{Hung2023PRR}, would only be possible at bias voltages of around $-3.6\,\text{V}$ and can thus be excluded by carefully selecting the bias voltage. From the diagram in \Figref{Fig2}d we can deduce that an energetically possible pathway to $D_1^+$ proceeds \textit{via} the neutral triplet state $T_1$. Neutralization of the molecule from $D_0^+$ can result in the formation of $T_1$\cite{Fatayer2021PRL,Peng2021Science}, which lies at around 1.2 eV\cite{Vincett1971JChemPhys,Nguyen2002JChemPhys,Chen2019PRL}. While the molecule is still in $T_1$, a subsequent charge-state transition can drive it into its charged excited state ($T_1 \rightarrow D_1^+$). The lifetimes of the two mediating states $T_1$ and $D_0^+$ lie within the same range, however, of the two only $D_0^+$ remains constant as a function of current and thus appears to dominate the anti-bunching width in X$^+$X$^+$. The bunching at larger $\Delta t$ relates to transitions out of the $D_0^+\rightarrow T_1\rightarrow D_1^+\rightarrow D_0^+$-loop, such as discharging from $D_0^+$ to $S_1$ or $S_0$ instead of $T_1$. Those transitions disable the X$^+$ emission for the time it takes to re-enter the loop and thus lead to blinking. Naturally, the mechanism leading to cross-bunching in X$^+$X involves all mediating states, \textit{i.e.} $S_0$, $T_1$, and $D_0^+$, such that the features in the X$^+$X correlation and their progression as a function of current cannot be easily assigned to individual transition rates. However, we find that anti-bunching occurs on the same timescale in both X$^+$X and XX, suggesting that also in X$^+$X the decrease in anti-bunching width is related to a decrease in $S_0$ lifetime. Bunching in X$^+$X occurs on comparable timescales, which may seem counter-intuitive at first glance since the lifetime of the state mediating the trion-exciton cascade, $D_0^+$, is shorter. This apparent discrepancy indicates that processes involving additional non-radiative loops after X$^+$ emission, such as $D_0^+\rightarrow T_1\rightarrow D_0^+$, also play a role in the cascaded emission. The correlation between the two emission channels is thus retained over several charge-state transitions.

\subsection*{Shifting the average population towards higher charge states}

\begin{figure}
\centering
    \includegraphics{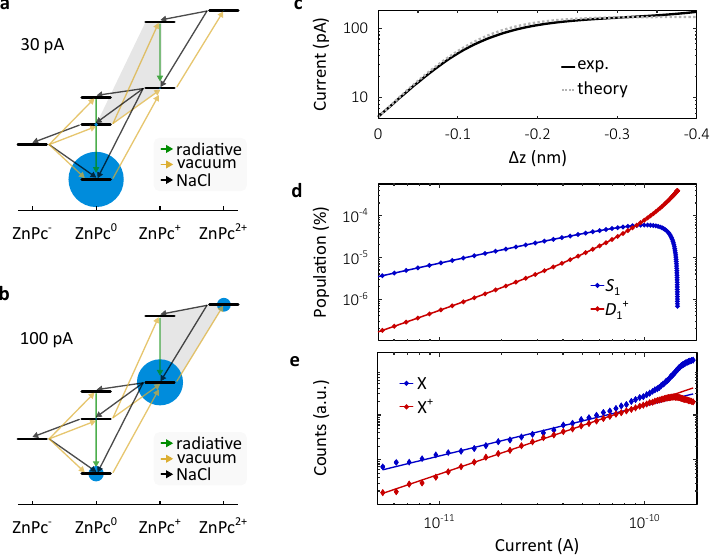}
    \caption{\textbf{Population dynamics as a function of tunnel current.} (a, b) Free energy diagram of the many-body states of ZnPc (see also \Figref{Fig2}d) together with the calculated population of the long-lived states at a setpoint of -3 V and (a) 30 pA and (b) 100 pA, respectively. The size of the blue circles indicates the relative population of the state. The grey shaded areas indicate the predominant pathway for trion formation. (c) Experimental (black line) and theoretically predicted (grey line) $I(\Delta z)$ atop ZnPc/4ML NaCl at -3 V. Dashed line: extrapolation of theoretical data. $\Delta z = 0\;\text{nm}$ corresponds to a setpoint of 5 pA at -3 V. (d) Calculated relative population of $S_1$ and $D_1^+$ as a function of tunnel current. (e) Experimental photon counts within an energy range of $1.87 - 1.91\,\text{eV}$ (blue) and $1.51 - 1.59\,\text{eV}$ (red) as a function of tunnel current, recorded at $V = -3\,\text{V}$. The blue and red curve correspond to the exciton and trion emission, respectively. The solid lines are fits to the data within the linear regime (in the double-logarithmic plot).}
    \label{Fig3}
\end{figure}

The gradual increase of the tip-mediated rates with respect to the sample-mediated rates shifts the ratio between charging and discharging probability. This affects the resonant charge transport through the molecule, as visualized by the flattening of the total tunnel current with decreasing $\Delta z$ (\Figref{Fig3}b), but also the average population of the different molecular states. \Figref{Fig3}a and b show the relative population (blue circles) of $S_0$, $T_1$, $D_0^+$ and $S_0^{2+}$ as extracted from the individual transition rates obtained from the fits for a calculated current of 30 pA and 100 pA. While in the low current limit ZnPc is neutral around 75\% of the time and doubly charged less than 1\%, at 100 pA the balance is strongly shifted toward ZnPc$^+$ (see Supporting Information Section S4 for a more detailed discussion). Consequently, transitions to higher positive charge states become increasingly likely as compared to transitions into neutral states. This has important consequences, especially for the correlation measurements. 
The bunching in X$^+$X$^+$ (\Figref{Fig2}b) is a result of $D_0^+\rightarrow S_0$ or $D_0^+\rightarrow S_1$ transitions leading to blinking in the X$^+$ emission. This blinking reduces as the population of $D_1^+$ \textit{via} higher charge states (\textit{e.g.} $S_0^{2+}$, grey shaded area in \Figref{Fig3}b) gradually replaces the one \textit{via} $T_1$ (grey shaded area in \Figref{Fig3}a). The same reasoning applies to the X$^+$X bunching peak (\Figref{Fig2}c), whose quenching with current can be directly associated to the lower probability of populating $S_1$ from $D_0^+$.

The detailed evolution of the excited states ($S_1$ and $D_1^+$) population with current can also be computed based on our rate equation model (\Figref{Fig3}d). Interestingly, one can get experimental insights into these parameters simply by studying the current dependence of the excitonic and trionic emission intensities (\Figref{Fig3}e).  At low currents, the simulations reveal that both  $S_1$ and $D_1^+$ populations follow a power-law dependence $\propto I_t^{\alpha}$, with $\alpha(\text{X}) < \alpha(\text{X}^+)$. As the current approaches saturation, both populations deviate from their low current power-laws, with $D_1^+$ increasing even further while $S_1$ decreases rapidly as expected from the gradual shift to higher charge states.
For currents up to $\approx 100\;\text{pA}$ we indeed find a qualitative agreement with the measured emission intensities (\Figref{Fig3}e). For higher currents, however, the emission intensities deviate from the trend in the excited state population. Up to $\approx 120\;\text{pA}$, this could be attributed to an increased coupling between the excitons and the picocavity plasmons which in turn increases the radiative decay probability \cite{Yang2020NatPhotonics,Imai-Imada2022Nature}. For even larger currents ($I >$ 120 pA), the X intensity saturates, an effect that can be attributed to the decrease in $S_1$ population (\Figref{Fig3}d). In contrast, the observed decrease in X$^+$ intensity in this current range cannot be accounted for based on the simulated population of $D_1$. This hints at two possible scenarios: 1) At sufficiently small $\Delta z$, charge transfer to the tip might compete with the radiative decay of one or both of the excited states and thus lead to luminescence quenching (\textit{e.g.} $S_1\rightarrow D_0^+$), as observed for example in photoluminescence experiments\cite{Imai-Imada2022Nature}. 2) Three- or more-electron processes, which play a role only at very high currents, might lead to the population of even higher-lying states (e.g. triply charged states). Both effects could lead to the quenching of the luminescence channels, and are not considered in our model.\\
Our results demonstrate that the electro-excitation of individual ZnPc leads to the generation of single photons from both its molecular exciton and trion. The cross-correlation of these emission channels reveals a correlated emission cascade, characterized by the radiative decay of the trion which is rapidly followed by the radiative decay of the exciton. The temporal dynamics of single photon emission are linked to charge-state transitions occurring between the molecule and the adjacent electrodes, namely the tip and metallic substrate. We elucidate the role of these transitions and the role of the intermediate electronic states in photon emission by experimentally manipulating the tip-mediated charging rate and employing a rate-equation-based theoretical model. With this model, we can describe the excitation dynamics up to tip-molecule distances where excited state quenching due to charge transfer from the tip becomes relevant. We identify the most relevant excitation paths leading to photon generation and highlight the importance of the triplet state $T_1$ as an intermediate step in the trion formation by sequential electron tunneling.  As such, our results not only provide a comprehensive picture of the electrically driven single and cascaded photon emission processes in a model QLS, but also demonstrate the potential of electrical pumping to address a variety of different excited states that can be utilized in non-classical light emission schemes, and lay out strategies for controlling their underlying photon statistics.

\section*{Acknowledgments}
We would like to thank S. Jiang, C. Schneider and S. Berciaud for fruitful discussions, and V. Speisser and H. Sumar for technical support. This project has received funding from the European Research Council (ERC) under the European Union's Horizon 2020 research and innovation program (grant agreement No 771850), the European Union's Horizon Europe research and innovation program under the Marie Sk\l{o}dowska-Curie grant agreement No 101059400, and the SNSF under the Postdoc.Mobility grant agreement No 206912. A. R. acknowledges funding from the Emmy Noether Programme of the the Deutsche Forschungsgemeinschaft (DFG, German Research Foundation) - 534367924. T. N. acknowledges the Lumina Quaeruntur fellowship of the Czech Academy of Sciences. Computational resources were supplied by the project "e-Infrastruktura CZ" (e-INFRA CZ LM2018140) supported by the Ministry of Education, Youth and Sports of the Czech Republic.\\
\noindent\textbf{Author contributions:} G.S. and K.K. designed the experiment. K.K. and A.R. performed the experiments. K.K., M.R., T.N., F.S. and G.S. discussed and analyzed the data. T.N. developed the theoretical model. All authors discussed the results and wrote the manuscript.

\bibliographystyle{naturemag}
\bibliography{HBT_ref}

\begin{thebibliography}{10}
\expandafter\ifx\csname url\endcsname\relax
  \def\url#1{\texttt{#1}}\fi
\expandafter\ifx\csname urlprefix\endcsname\relax\def\urlprefix{URL }\fi
\providecommand{\bibinfo}[2]{#2}
\providecommand{\eprint}[2][]{\url{#2}}

\bibitem{Kurtsiefer2000PRL}
\bibinfo{author}{Kurtsiefer, C.}, \bibinfo{author}{Mayer, S.}, \bibinfo{author}{Zarda, P.} \& \bibinfo{author}{Weinfurter, H.}
\newblock \bibinfo{title}{Stable solid-state source of single photons}.
\newblock \emph{\bibinfo{journal}{Phys. Rev. Lett.}} \textbf{\bibinfo{volume}{85}}, \bibinfo{pages}{290--293} (\bibinfo{year}{2000}).

\bibitem{Rogers2014NatCommun}
\bibinfo{author}{Rogers, L.~J.} \emph{et~al.}
\newblock \bibinfo{title}{Multiple intrinsically identical single-photon emitters in the solid state}.
\newblock \emph{\bibinfo{journal}{Nat. Commun.}} \textbf{\bibinfo{volume}{5}}, \bibinfo{pages}{4739} (\bibinfo{year}{2014}).

\bibitem{Bradac2019NatCommun}
\bibinfo{author}{Bradac, C.}, \bibinfo{author}{Gao, W.}, \bibinfo{author}{Forneris, J.}, \bibinfo{author}{Trusheim, M.~E.} \& \bibinfo{author}{Aharonovich, I.}
\newblock \bibinfo{title}{Quantum nanophotonics with group iv defects in diamond}.
\newblock \emph{\bibinfo{journal}{Nat. Commun.}} \textbf{\bibinfo{volume}{10}}, \bibinfo{pages}{5625} (\bibinfo{year}{2019}).

\bibitem{Higginbottom2022Nature}
\bibinfo{author}{Higginbottom, D.~B.} \emph{et~al.}
\newblock \bibinfo{title}{Optical observation of single spins in silicon}.
\newblock \emph{\bibinfo{journal}{Nature}} \textbf{\bibinfo{volume}{607}}, \bibinfo{pages}{266--270} (\bibinfo{year}{2022}).

\bibitem{Prabhu2023NatCommun}
\bibinfo{author}{Prabhu, M.} \emph{et~al.}
\newblock \bibinfo{title}{Individually addressable and spectrally programmable artificial atoms in silicon photonics}.
\newblock \emph{\bibinfo{journal}{Nat. Commun.}} \textbf{\bibinfo{volume}{14}} (\bibinfo{year}{2023}).

\bibitem{Michler2000Science}
\bibinfo{author}{Michler, P.} \emph{et~al.}
\newblock \bibinfo{title}{A quantum dot single-photon turnstile device}.
\newblock \emph{\bibinfo{journal}{Science}} \textbf{\bibinfo{volume}{290}}, \bibinfo{pages}{2282--2285} (\bibinfo{year}{2000}).

\bibitem{Garcia2021Science}
\bibinfo{author}{Garc\'{i}a~de Arquer, F.~P.} \emph{et~al.}
\newblock \bibinfo{title}{Semiconductor quantum dots: Technological progress and future challenges}.
\newblock \emph{\bibinfo{journal}{Science}} \textbf{\bibinfo{volume}{373}}, \bibinfo{pages}{eaaz8541} (\bibinfo{year}{2021}).

\bibitem{Brunel1999PRL}
\bibinfo{author}{Brunel, C.}, \bibinfo{author}{Lounis, B.}, \bibinfo{author}{Tamarat, P.} \& \bibinfo{author}{Orrit, M.}
\newblock \bibinfo{title}{Triggered source of single photons based on controlled single molecule fluorescence}.
\newblock \emph{\bibinfo{journal}{Phys. Rev. Lett.}} \textbf{\bibinfo{volume}{83}}, \bibinfo{pages}{2722--2725} (\bibinfo{year}{1999}).

\bibitem{Lounis2000Nature}
\bibinfo{author}{Lounis, B.} \& \bibinfo{author}{Moerner, W.}
\newblock \bibinfo{title}{Single photons on demand from a single molecule at room temperature}.
\newblock \emph{\bibinfo{journal}{Nature}} \textbf{\bibinfo{volume}{407}}, \bibinfo{pages}{491--493} (\bibinfo{year}{2000}).

\bibitem{Wang2019NatPhys}
\bibinfo{author}{Wang, D.} \emph{et~al.}
\newblock \bibinfo{title}{Turning a molecule into a coherent two-level quantum system}.
\newblock \emph{\bibinfo{journal}{Nat. Phys.}} \textbf{\bibinfo{volume}{15}}, \bibinfo{pages}{483--489} (\bibinfo{year}{2019}).

\bibitem{Murtaza2023OptExpress}
\bibinfo{author}{Murtaza, G.} \emph{et~al.}
\newblock \bibinfo{title}{Efficient room-temperature molecular single-photon sources for quantum key distribution}.
\newblock \emph{\bibinfo{journal}{Opt. Express}} \textbf{\bibinfo{volume}{31}}, \bibinfo{pages}{9437--9447} (\bibinfo{year}{2023}).

\bibitem{Toninelli2021NatMater}
\bibinfo{author}{Toninelli, C.} \emph{et~al.}
\newblock \bibinfo{title}{Single organic molecules for photonic quantum technologies}.
\newblock \emph{\bibinfo{journal}{Nat. Mater.}} \textbf{\bibinfo{volume}{20}}, \bibinfo{pages}{1615--1628} (\bibinfo{year}{2021}).

\bibitem{Waks2002Nature}
\bibinfo{author}{Waks, E.} \emph{et~al.}
\newblock \bibinfo{title}{Quantum cryptography with a photon turnstile}.
\newblock \emph{\bibinfo{journal}{Nature}} \textbf{\bibinfo{volume}{420}}, \bibinfo{pages}{762} (\bibinfo{year}{2002}).

\bibitem{Beveratos2002PRL}
\bibinfo{author}{Beveratos, A.} \emph{et~al.}
\newblock \bibinfo{title}{Single photon quantum cryptography}.
\newblock \emph{\bibinfo{journal}{Phys. Rev. Lett.}} \textbf{\bibinfo{volume}{89}}, \bibinfo{pages}{187901} (\bibinfo{year}{2002}).

\bibitem{OBrien2007Science}
\bibinfo{author}{O'Brien, J.~L.}
\newblock \bibinfo{title}{Optical quantum computing}.
\newblock \emph{\bibinfo{journal}{Science}} \textbf{\bibinfo{volume}{318}}, \bibinfo{pages}{1567--1570} (\bibinfo{year}{2007}).

\bibitem{Kimble2008Nature}
\bibinfo{author}{Kimble, H.~J.}
\newblock \bibinfo{title}{The quantum internet}.
\newblock \emph{\bibinfo{journal}{Nature}} \textbf{\bibinfo{volume}{453}}, \bibinfo{pages}{1023--1030} (\bibinfo{year}{2008}).

\bibitem{Yuan2002Science}
\bibinfo{author}{Yuan, Z.} \emph{et~al.}
\newblock \bibinfo{title}{Electrically driven single-photon source}.
\newblock \emph{\bibinfo{journal}{Science}} \textbf{\bibinfo{volume}{295}}, \bibinfo{pages}{102--105} (\bibinfo{year}{2002}).

\bibitem{Salter2010Nature}
\bibinfo{author}{Salter, C.~L.} \emph{et~al.}
\newblock \bibinfo{title}{An entangled-light-emitting diode}.
\newblock \emph{\bibinfo{journal}{Nature}} \textbf{\bibinfo{volume}{465}}, \bibinfo{pages}{594--597} (\bibinfo{year}{2010}).

\bibitem{Nothaft2012NatCommun}
\bibinfo{author}{Nothaft, M.}, \bibinfo{author}{H{\"{o}}hla, S.}, \bibinfo{author}{Jelezko, F.}, \bibinfo{author}{Fr{\"{u}}hauf, N.} \& \bibinfo{author}{Wrachtrup, J.}
\newblock \bibinfo{title}{Electrically driven photon antibunching from a single molecule at room temperature}.
\newblock \emph{\bibinfo{journal}{Nat. Commun.}} \textbf{\bibinfo{volume}{3}} (\bibinfo{year}{2012}).

\bibitem{Mizuochi2012NatPhoton}
\bibinfo{author}{Mizuochi, N.} \emph{et~al.}
\newblock \bibinfo{title}{Electrically driven single-photon source at room temperature in diamond}.
\newblock \emph{\bibinfo{journal}{Nat. Photon.}} \textbf{\bibinfo{volume}{6}}, \bibinfo{pages}{299--303} (\bibinfo{year}{2012}).

\bibitem{Bentham2016ApplPhysLett}
\bibinfo{author}{Bentham, C.} \emph{et~al.}
\newblock \bibinfo{title}{Single-photon electroluminescence for on-chip quantum networks}.
\newblock \emph{\bibinfo{journal}{Appl. Phys. Lett.}} \textbf{\bibinfo{volume}{109}}, \bibinfo{pages}{161101} (\bibinfo{year}{2016}).

\bibitem{Lin2017NatCommun}
\bibinfo{author}{Lin, X.} \emph{et~al.}
\newblock \bibinfo{title}{Electrically-driven single-photon sources based on colloidal quantum dots with near-optimal antibunching at room temperature}.
\newblock \emph{\bibinfo{journal}{Nat. Commun.}} \textbf{\bibinfo{volume}{8}}, \bibinfo{pages}{1132} (\bibinfo{year}{2017}).

\bibitem{Fischer2017NaturePhys}
\bibinfo{author}{Fischer, K.~A.} \emph{et~al.}
\newblock \bibinfo{title}{Signatures of two-photon pulses from a quantum two-level system}.
\newblock \emph{\bibinfo{journal}{Nat. Phys.}} \textbf{\bibinfo{volume}{13}}, \bibinfo{pages}{649--654} (\bibinfo{year}{2017}).

\bibitem{Aharonovich2016NatPhoton}
\bibinfo{author}{Aharonovich, I.}, \bibinfo{author}{Englund, D.} \& \bibinfo{author}{Toth, M.}
\newblock \bibinfo{title}{Solid-state single-photon emitters}.
\newblock \emph{\bibinfo{journal}{Nat. Photon.}} \textbf{\bibinfo{volume}{10}}, \bibinfo{pages}{631--641} (\bibinfo{year}{2016}).

\bibitem{Jiang2023PRL}
\bibinfo{author}{Jiang, S.} \emph{et~al.}
\newblock \bibinfo{title}{Many-body description of stm-induced fluorescence of charged molecules}.
\newblock \emph{\bibinfo{journal}{Phys. Rev. Lett.}} \textbf{\bibinfo{volume}{130}}, \bibinfo{pages}{126202} (\bibinfo{year}{2023}).

\bibitem{Qiu2003Science}
\bibinfo{author}{Qiu, X.~H.}, \bibinfo{author}{Nazin, G.~V.} \& \bibinfo{author}{Ho, W.}
\newblock \bibinfo{title}{Vibrationally {Resolved} {Fluorescence} {Excited} with {Submolecular} {Precision}}.
\newblock \emph{\bibinfo{journal}{Science}} \textbf{\bibinfo{volume}{299}}, \bibinfo{pages}{542--546} (\bibinfo{year}{2003}).

\bibitem{Zhang2016Nature}
\bibinfo{author}{Zhang, Y.} \emph{et~al.}
\newblock \bibinfo{title}{Visualizing coherent intermolecular dipole-dipole coupling in real space}.
\newblock \emph{\bibinfo{journal}{Nature}} \textbf{\bibinfo{volume}{531}}, \bibinfo{pages}{623--627} (\bibinfo{year}{2016}).

\bibitem{Imada2016Nature}
\bibinfo{author}{Imada, H.} \emph{et~al.}
\newblock \bibinfo{title}{Real-space investigation of energy transfer in heterogeneous molecular dimers}.
\newblock \emph{\bibinfo{journal}{Nature}} \textbf{\bibinfo{volume}{538}}, \bibinfo{pages}{364--367} (\bibinfo{year}{2016}).

\bibitem{Doppagne2017PRL}
\bibinfo{author}{Doppagne, B.} \emph{et~al.}
\newblock \bibinfo{title}{Vibronic spectroscopy with submolecular resolution from stm-induced electroluminescence}.
\newblock \emph{\bibinfo{journal}{Phys. Rev. Lett.}} \textbf{\bibinfo{volume}{118}}, \bibinfo{pages}{127401} (\bibinfo{year}{2017}).

\bibitem{Brown1956Nature}
\bibinfo{author}{Hanbury~Brown, R.} \& \bibinfo{author}{Twiss, R.~Q.}
\newblock \bibinfo{title}{Correlation between photons in two coherent beams of light}.
\newblock \emph{\bibinfo{journal}{Nature}} \textbf{\bibinfo{volume}{177}}, \bibinfo{pages}{27--29} (\bibinfo{year}{1956}).

\bibitem{Perronet2006EPL}
\bibinfo{author}{Perronet, K.}, \bibinfo{author}{Schull, G.}, \bibinfo{author}{Raimond, P.} \& \bibinfo{author}{Charra, F.}
\newblock \bibinfo{title}{Single-molecule fluctuations in a tunnel junction: A study by scanning-tunnelling-microscopy–induced luminescence}.
\newblock \emph{\bibinfo{journal}{EPL}} \textbf{\bibinfo{volume}{74}}, \bibinfo{pages}{313} (\bibinfo{year}{2006}).

\bibitem{Merino2015NatCommun}
\bibinfo{author}{Merino, P.}, \bibinfo{author}{Gro{\ss}e, C.}, \bibinfo{author}{Ros{\l}awska, A.}, \bibinfo{author}{Kuhnke, K.} \& \bibinfo{author}{Kern, K.}
\newblock \bibinfo{title}{Exciton dynamics of c60-based single-photon emitters explored by hanbury brown--twiss scanning tunnelling microscopy}.
\newblock \emph{\bibinfo{journal}{Nat. Commun.}} \textbf{\bibinfo{volume}{6}}, \bibinfo{pages}{8461} (\bibinfo{year}{2015}).

\bibitem{Zhang2017NatCommun}
\bibinfo{author}{Zhang, L.} \emph{et~al.}
\newblock \bibinfo{title}{Electrically driven single-photon emission from an isolated single molecule}.
\newblock \emph{\bibinfo{journal}{Nat. Commun.}} \textbf{\bibinfo{volume}{8}}, \bibinfo{pages}{580} (\bibinfo{year}{2017}).

\bibitem{Luo2019PRL}
\bibinfo{author}{Luo, Y.} \emph{et~al.}
\newblock \bibinfo{title}{Electrically driven single-photon superradiance from molecular chains in a plasmonic nanocavity}.
\newblock \emph{\bibinfo{journal}{Phys. Rev. Lett.}} \textbf{\bibinfo{volume}{122}}, \bibinfo{pages}{233901} (\bibinfo{year}{2019}).

\bibitem{Roslawska2020ACSNano}
\bibinfo{author}{Ros{\l{}}awska, A.} \emph{et~al.}
\newblock \bibinfo{title}{Atomic-scale dynamics probed by photon correlations}.
\newblock \emph{\bibinfo{journal}{ACS Nano}} \textbf{\bibinfo{volume}{14}}, \bibinfo{pages}{6366--6375} (\bibinfo{year}{2020}).

\bibitem{Yersin2004Book}
\bibinfo{author}{Yersin, H.}
\newblock \emph{\bibinfo{title}{Triplet Emitters for OLED Applications. Mechanisms of Exciton Trapping and Control of Emission Properties}}, \bibinfo{pages}{1--26} (\bibinfo{publisher}{Springer Berlin Heidelberg}, \bibinfo{address}{Berlin, Heidelberg}, \bibinfo{year}{2004}).

\bibitem{Doppagne2018Science}
\bibinfo{author}{Doppagne, B.} \emph{et~al.}
\newblock \bibinfo{title}{Electrofluorochromism at the single-molecule level}.
\newblock \emph{\bibinfo{journal}{Science}} \textbf{\bibinfo{volume}{361}}, \bibinfo{pages}{251--255} (\bibinfo{year}{2018}).

\bibitem{Dolezal2021ACSNano}
\bibinfo{author}{Dole{\v{z}}al, J.}, \bibinfo{author}{Canola, S.}, \bibinfo{author}{Merino, P.} \& \bibinfo{author}{{\v{S}}vec, M.}
\newblock \bibinfo{title}{Exciton-{Trion} {Conversion} {Dynamics} in a {Single} {Molecule}}.
\newblock \emph{\bibinfo{journal}{ACS Nano}} \textbf{\bibinfo{volume}{15}}, \bibinfo{pages}{7694--7699} (\bibinfo{year}{2021}).

\bibitem{Hung2023PRR}
\bibinfo{author}{Hung, T.-C.} \emph{et~al.}
\newblock \bibinfo{title}{Bipolar single-molecule electroluminescence and electrofluorochromism}.
\newblock \emph{\bibinfo{journal}{Phys. Rev. Res.}} \textbf{\bibinfo{volume}{5}}, \bibinfo{pages}{033027} (\bibinfo{year}{2023}).

\bibitem{Roslawska2018NanoLett}
\bibinfo{author}{Ros{\l{}}awska, A.} \emph{et~al.}
\newblock \bibinfo{title}{Single {Charge} and {Exciton} {Dynamics} {Probed} by {Molecular}-{Scale}-{Induced} {Electroluminescence}}.
\newblock \emph{\bibinfo{journal}{Nano Lett.}} \textbf{\bibinfo{volume}{18}}, \bibinfo{pages}{4001--4007} (\bibinfo{year}{2018}).

\bibitem{Miwa2019NanoLett}
\bibinfo{author}{Miwa, K.} \emph{et~al.}
\newblock \bibinfo{title}{Many-{Body} {State} {Description} of {Single}-{Molecule} {Electroluminescence} {Driven} by a {Scanning} {Tunneling} {Microscope}}.
\newblock \emph{\bibinfo{journal}{Nano Lett.}} \textbf{\bibinfo{volume}{19}}, \bibinfo{pages}{2803--2811} (\bibinfo{year}{2019}).

\bibitem{Beveratos2001PRA}
\bibinfo{author}{Beveratos, A.}, \bibinfo{author}{Brouri, R.}, \bibinfo{author}{Gacoin, T.}, \bibinfo{author}{Poizat, J.-P.} \& \bibinfo{author}{Grangier, P.}
\newblock \bibinfo{title}{Nonclassical radiation from diamond nanocrystals}.
\newblock \emph{\bibinfo{journal}{Phys. Rev. A}} \textbf{\bibinfo{volume}{64}}, \bibinfo{pages}{061802} (\bibinfo{year}{2001}).

\bibitem{Fishman2023PRX}
\bibinfo{author}{Fishman, R.~E.}, \bibinfo{author}{Patel, R.~N.}, \bibinfo{author}{Hopper, D.~A.}, \bibinfo{author}{Huang, T.-Y.} \& \bibinfo{author}{Bassett, L.~C.}
\newblock \bibinfo{title}{Photon-emission-correlation spectroscopy as an analytical tool for solid-state quantum defects}.
\newblock \emph{\bibinfo{journal}{PRX Quantum}} \textbf{\bibinfo{volume}{4}}, \bibinfo{pages}{010202} (\bibinfo{year}{2023}).

\bibitem{Roslawska2022PRX}
\bibinfo{author}{Ros{\l}awska, A.} \emph{et~al.}
\newblock \bibinfo{title}{Mapping {Lamb}, {Stark}, and {Purcell} {Effects} at a {Chromophore}-{Picocavity} {Junction} with {Hyper}-{Resolved} {Fluorescence} {Microscopy}}.
\newblock \emph{\bibinfo{journal}{Phys. Rev. X}} \textbf{\bibinfo{volume}{12}}, \bibinfo{pages}{011012} (\bibinfo{year}{2022}).

\bibitem{Dolezal2024NanoLett}
\bibinfo{author}{Dole\v{z}al, J.}, \bibinfo{author}{Sagwal, A.}, \bibinfo{author}{de~Campos~Ferreira, R.~C.} \& \bibinfo{author}{\v{S}vec, M.}
\newblock \bibinfo{title}{Single-molecule time-resolved spectroscopy in a tunable stm nanocavity}.
\newblock \emph{\bibinfo{journal}{Nano Lett.}} \textbf{\bibinfo{volume}{24}}, \bibinfo{pages}{1629--1634} (\bibinfo{year}{2024}).

\bibitem{Kaiser2023NatCommun}
\bibinfo{author}{Kaiser, K.}, \bibinfo{author}{Lieske, L.-A.}, \bibinfo{author}{Repp, J.} \& \bibinfo{author}{Gross, L.}
\newblock \bibinfo{title}{Charge-state lifetimes of single molecules on few monolayers of {NaCl}}.
\newblock \emph{\bibinfo{journal}{Nat. Commun.}} \textbf{\bibinfo{volume}{14}} (\bibinfo{year}{2023}).

\bibitem{Steurer2014ApplPhysLett}
\bibinfo{author}{Steurer, W.}, \bibinfo{author}{Gross, L.} \& \bibinfo{author}{Meyer, G.}
\newblock \bibinfo{title}{Local thickness determination of thin insulator films via localized states}.
\newblock \emph{\bibinfo{journal}{Appl. Phys. Lett.}} \textbf{\bibinfo{volume}{104}}, \bibinfo{pages}{231606} (\bibinfo{year}{2014}).

\bibitem{Draebenstedt1999PRB}
\bibinfo{author}{Dr\"abenstedt, A.} \emph{et~al.}
\newblock \bibinfo{title}{Low-temperature microscopy and spectroscopy on single defect centers in diamond}.
\newblock \emph{\bibinfo{journal}{Phys. Rev. B}} \textbf{\bibinfo{volume}{60}}, \bibinfo{pages}{11503--11508} (\bibinfo{year}{1999}).

\bibitem{Schroeder2021ACSNano}
\bibinfo{author}{Schr\"{o}der, T.} \emph{et~al.}
\newblock \bibinfo{title}{How blinking affects photon correlations in multichromophoric nanoparticles}.
\newblock \emph{\bibinfo{journal}{ACS Nano}} \textbf{\bibinfo{volume}{15}}, \bibinfo{pages}{18037--18047} (\bibinfo{year}{2021}).

\bibitem{Aspect1980PRL}
\bibinfo{author}{Aspect, A.}, \bibinfo{author}{Roger, G.}, \bibinfo{author}{Reynaud, S.}, \bibinfo{author}{Dalibard, J.} \& \bibinfo{author}{Cohen-Tannoudji, C.}
\newblock \bibinfo{title}{Time correlations between the two sidebands of the resonance fluorescence triplet}.
\newblock \emph{\bibinfo{journal}{Phys. Rev. Lett.}} \textbf{\bibinfo{volume}{45}}, \bibinfo{pages}{617--620} (\bibinfo{year}{1980}).

\bibitem{Moreau2001PRL}
\bibinfo{author}{Moreau, E.} \emph{et~al.}
\newblock \bibinfo{title}{Quantum cascade of photons in semiconductor quantum dots}.
\newblock \emph{\bibinfo{journal}{Phys. Rev. Lett.}} \textbf{\bibinfo{volume}{87}}, \bibinfo{pages}{183601} (\bibinfo{year}{2001}).

\bibitem{He2016NatCommun}
\bibinfo{author}{He, Y.-M.} \emph{et~al.}
\newblock \bibinfo{title}{Cascaded emission of single photons from the biexciton in monolayered wse2}.
\newblock \emph{\bibinfo{journal}{Nat. Commun.}} \textbf{\bibinfo{volume}{7}}, \bibinfo{pages}{13409} (\bibinfo{year}{2016}).

\bibitem{Fatayer2021PRL}
\bibinfo{author}{Fatayer, S.} \emph{et~al.}
\newblock \bibinfo{title}{Probing molecular excited states by atomic force microscopy}.
\newblock \emph{\bibinfo{journal}{Phys. Rev. Lett.}} \textbf{\bibinfo{volume}{126}}, \bibinfo{pages}{176801} (\bibinfo{year}{2021}).

\bibitem{Peng2021Science}
\bibinfo{author}{Peng, J.} \emph{et~al.}
\newblock \bibinfo{title}{Atomically resolved single-molecule triplet quenching}.
\newblock \emph{\bibinfo{journal}{Science}} \textbf{\bibinfo{volume}{373}}, \bibinfo{pages}{452--456} (\bibinfo{year}{2021}).

\bibitem{Vincett1971JChemPhys}
\bibinfo{author}{Vincett, P.~S.}, \bibinfo{author}{Voigt, E.~M.} \& \bibinfo{author}{Rieckhoff, K.~E.}
\newblock \bibinfo{title}{Phosphorescence and fluorescence of phthalocyanines}.
\newblock \emph{\bibinfo{journal}{J. Chem. Phys.}} \textbf{\bibinfo{volume}{55}}, \bibinfo{pages}{4131--4140} (\bibinfo{year}{1971}).

\bibitem{Nguyen2002JChemPhys}
\bibinfo{author}{Nguyen, K.~A.}, \bibinfo{author}{Kennel, J.} \& \bibinfo{author}{Pachter, R.}
\newblock \bibinfo{title}{A density functional theory study of phosphorescence and triplet–triplet absorption for nonlinear absorption chromophores}.
\newblock \emph{\bibinfo{journal}{J. Chem. Phys.}} \textbf{\bibinfo{volume}{117}}, \bibinfo{pages}{7128--7136} (\bibinfo{year}{2002}).

\bibitem{Chen2019PRL}
\bibinfo{author}{Chen, G.} \emph{et~al.}
\newblock \bibinfo{title}{Spin-triplet-mediated up-conversion and crossover behavior in single-molecule electroluminescence}.
\newblock \emph{\bibinfo{journal}{Phys. Rev. Lett.}} \textbf{\bibinfo{volume}{122}}, \bibinfo{pages}{177401} (\bibinfo{year}{2019}).

\bibitem{Yang2020NatPhotonics}
\bibinfo{author}{Yang, B.} \emph{et~al.}
\newblock \bibinfo{title}{Sub-nanometre resolution in single-molecule photoluminescence imaging}.
\newblock \emph{\bibinfo{journal}{Nat. Photonics}} \textbf{\bibinfo{volume}{14}}, \bibinfo{pages}{693--699} (\bibinfo{year}{2020}).

\bibitem{Imai-Imada2022Nature}
\bibinfo{author}{Imai-Imada, M.} \emph{et~al.}
\newblock \bibinfo{title}{Orbital-resolved visualization of single-molecule photocurrent channels}.
\newblock \emph{\bibinfo{journal}{Nature}} \textbf{\bibinfo{volume}{603}}, \bibinfo{pages}{829--834} (\bibinfo{year}{2022}).

\end{thebibliography}



\end{document}


\title{Supporting Information\\Electrically driven cascaded photon-emission in a single molecule}

\author
{Katharina Kaiser,$^{1,2\ast}$ Anna Ros{\l}awska,$^{1,3}$ Michelangelo Romeo,$^{1}$\\ 
Fabrice Scheurer,$^{1}$ Tom{\'a}\v{s} Neuman,$^{4,5\ast}$ Guillaume Schull$^{1\ast}$\\
\normalsize{$^{1}$Universit\'{e} de Strasbourg, IPCMS, CNRS, UMR 7504, Strasbourg, France}\\
\normalsize{$^{2}$4th Physical Institute -- Solids and Nanostructures, Georg-August-Universit\"{a}t G\"{o}ttingen}\\
\normalsize{G\"{o}ttingen, 37077, Germany}\\
\normalsize{$^{3}$Max Planck Institute for Solid State Research, Stuttgart, 70569, Germany}\\
\normalsize{$^{4}$Universit\'{e} Paris-Saclay, Institut des Sciences Mol\'{e}culaires d'Orsay (ISMO)}\\
\normalsize{CNRS, UMR 8214, Orsay, 91405, France}\\
\normalsize{$^{5}$Institute of Physics, Czech Academy of Sciences}\\
\normalsize{Cukrovarnick\'{a} 10, Prague, 16200, Czech Republic}\\
\normalsize{$^\ast$To whom correspondence should be addressed:}\\
\normalsize{E-mail: katharina.kaiser@uni-goettingen.de; neuman@fzu.cz; schull@unistra.fr.}
}

\sloppy

\maketitle

\newpage
\tableofcontents

\newpage
\noindent

\section*{S1 -- Experimental methods}

\addcontentsline{toc}{subsection}{Sample preparation}
\noindent \textbf{Sample preparation}\\
The experiments were performed on isolated zinc-phthalocyanine (ZnPc) molecules adsorbed on 4 monolayers (ML) NaCl on a Ag(111)-substrate. To that end, the Ag single-crystal was cleaned \textit{in situ} by repeated Ar$^+$-ion sputtering and annealing cycles. NaCl was thermally sublimed onto the clean Ag(111) surface kept at room temperature, followed by mild annealing, resulting in the formation of NaCl-islands of 3-5 ML thickness. ZnPc (Sigma-Aldrich) was thermally sublimed from a crucible onto the cold ($T\approx 7\;\text{K}$) sample, yielding a sub-ML coverage of well-dispersed molecules. We used electrochemically etched Ag-tips, that were optimized for STM-imaging and STML by indentations into the Ag-crystal and voltage pulses.
\vspace{1cm}

\begin{figure}
    \centering
    \includegraphics[width = 0.8\textwidth]{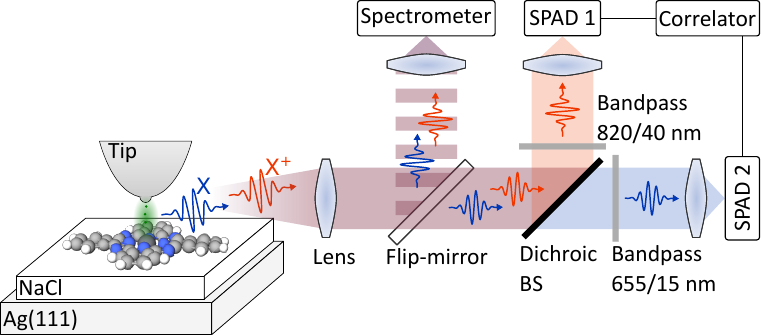}
    \caption{\textbf{Schematic of the STML setup.} The flip-mirror is used to change between recording STML spectra and correlation measurements with the HBT setup. BS: Beam splitter; SPAD: Single-photon avalanche diode. The schematic shows the setup for detecting the X$^+$X correlation. For XX and X$^+$X$^+$, the dichroic BS is replaced by a 50:50 BS, and the bandpass filters are chosen to fit the respective emission lines on both sides.}
    \label{FigS1}
\end{figure}

\addcontentsline{toc}{subsection}{STM and optical setup}
\noindent \textbf{STM and optical setup}
The experiments were performed in a low-temperature ($T \approx 6\,\text{K}$) ultra-high vacuum (UHV) STM with optical access (Unisoku USM1400). Two movable lenses with a numerical aperture of 0.55 each are located in the STM head. A schematic overview of the optical setup is given in \Figref{FigS1}. The light emitted from the junction is collected and collimated by the \textit{in-situ} lenses and is guided either towards a Hanbury Brown-Twiss (HBT) setup or, by placing a flippable mirror into the beam path, a spectrometer with a liquid N$_2$-cooled CCD camera (Princeton Instruments), both positioned outside the vacuum chamber. Note that here, we only show the detection using one of the lenses. Alternatively, one of the arms of the HBT setup can be moved to the side of the second lens. The HBT interferometry was performed in two different configurations with two single-photon avalanche diodes (SPAD, Excelitas or MPD) being positioned either both on one side of the chamber, \textit{i.e.}, using only one of the \textit{in-situ} lenses, or one on each side, \textit{i.e.}, using both \textit{in-situ} lenses. In both configurations, the SPADs were connected to a Time-Correlated Single Photon Counting system (TCSPC, PicoHarp 300) and to the Nanonis electronics. The time resolution of the HBT measurement is ultimately limited by the time resolution of the utilized SPADs. The time resolution for the MPD is 35 ps, and 250 ps for Excelitas, which in turn offers an increased detection efficiency as compared to the MPD SPAD.\\

For the detection on one side, the beam path was split using a 50:50 beamsplitter or a dichroic mirror (transmitting light with $\lambda > 785\,\text{nm}$ and reflecting light with $\lambda < 785\,\text{nm}$, as shown in \Figref{FigS1}), depending on which emission line was recorded. To select the photons emitted from the X and X$^+$ transitions, respectively, we used bandpass filters ($655\pm 7.5\,\text{nm}$ or a combination of a $700\,\text{nm}$ shortpass and a $650\,\text{nm}$ longpass for the neutral exciton, $800\pm20\,\text{nm}$ for the trion). The (filtered) signal was subsequently focused onto the SPADs.
\vspace{1cm}

\addcontentsline{toc}{subsection}{Hanbury Brown-Twiss measurements}
\noindent\textbf{Hanbury Brown-Twiss measurements}\\
In HBT interferometry, two SPADs are used to record photons emitted from the junction. The signal is forwarded to a TCSPC, tracking the corresponding arrival times. The two channels are offset in the PicoHarp software to allow the detection of negative time delays $\Delta t$ of two consecutive photons. $\Delta t$ and the (non-normalized) second order correlation function $g^{(2)}(\Delta t)$ can be measured in two ways: 1. In the \textit{histogram mode}, one detector is selected as 'start'. A photon detected on this channel starts the timer of the TCSPC, the next photon detected on the second channel stops it. A histogram of these delay times then yields $g^{(2)}(\Delta t)$. In this mode, only $\Delta t$ but not the absolute arrival time is recorded and thus, the time resolution of the measurement (binning) needs to be set \textit{a priori} and cannot be changed afterwards. 2. In the \textit{T2-mode}, the detected intensities $I_{\alpha,\beta}$ of X and X$^+$, respectively, with $\alpha$ and $\beta$ in $\{\text{X}, \text{X}^+\}$, are recorded in full. This allows extracting $\Delta t$ and setting the binning for the histogram of $g^{(2)}(\Delta t)$ \textit{a posteriori}. The binning for the $g^{(2)}(\Delta t)$ curves shown in the main text and Supporting Information has been set to 128 ps. $g^{(2)}(\Delta t)$ is calculated according to

\begin{align*}
    g^{(2)}(\Delta t) = \frac{\braket{I_{\alpha}(t)I_{\beta}(t+\Delta t)}}{\braket{{I_{\alpha}(t)}}\braket{I_{\beta}(t)}}.
\end{align*}

The obtained $g^{(2)}(\Delta t)$ is normalized by $g^{(2)}(\Delta t)$ at  $\Delta t\rightarrow \infty$.\\

For correlation measurements on ZnPc, the tip is initially positioned atop a maximum of the orbital density (HOMO in this case). The atom tracking feature of the Nanonis software is then set to the photon count recorded by one of the SPAD detectors to keep it at maximum, typically the one recording the emission from the trion. Like this, the photon count in one of the channels can be maximized. Measurement of $g^{(2)}(\Delta t)$ atop different positions of the molecule (\textit{i.e.}, maximized for trionic or excitonic emission) does not yield different results (see also section S2). This approach merely facilitates a drastic reduction in measurement time and ensures that the tip stays atop the molecule throughout a measurement. Nevertheless, typical recording times per $g^{(2)}(\Delta t)$ measurement ranged from a few up to around 20 hours. In some occasions, the molecule was picked up by the tip during a measurement. In this case, we typically interrupted the recording of the TCSPC, re-prepared the tip (by soft indentations of the tip into the bare metallic substrate), moved to another molecule, and continued the data acquisition of the TCSPC. Thus, some of the shown histograms were recorded atop several molecules.
\vspace{1cm}

\begin{figure}[ht]
    \centering
    \includegraphics{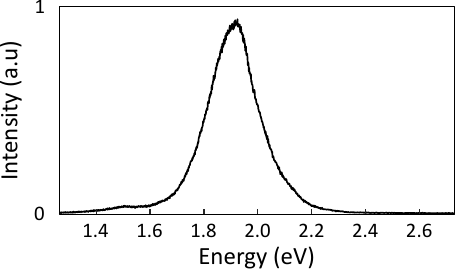}
    \caption{\textbf{Plasmonic spectra of the tip.} Typical STML spectrum recorded on the bare Ag(111) surface at 3 V and 500 pA with a grating of $150\;\text{grooves/mm}$, showing the plasmonic electroluminescence.}
    \label{FigS2}
\end{figure}

\addcontentsline{toc}{subsection}{Characterization of the plasmonic response of the tip}
\noindent\textbf{Characterization of the plasmonic response of the tip}\\
In order to optimize the luminescence yield, the plasmonic response of the tip is conditioned by voltage pulses and careful indentations into the bare metallic substrate. \Figref{FigS2} shows a typical plasmonic spectrum recorded atop the bare Ag(111) surface after optimization.
\newpage

\newpage

\section*{S2 -- Additional correlation data}
\begin{figure}[!ht]
\centering
\includegraphics{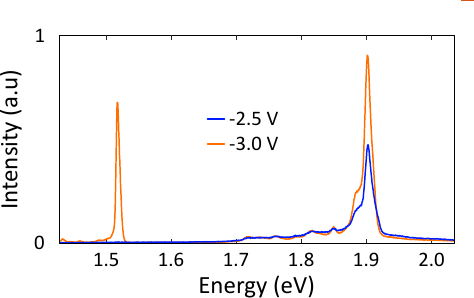}
\caption{\textbf{Voltage dependent STML spectra.} STML spectra recorded at setpoints of 130\,pA and -2.5\,V (blue) and -3\,V (orange), respectively. Integration time 60\,s (-2.5\,V, blue curve) and 20\,s (-3\,V, orange curve), 300 grooves/mm.}
\label{FigS3_1}
\end{figure}

\begin{figure}[!ht]
    \centering
    \includegraphics[width = 0.75\textwidth]{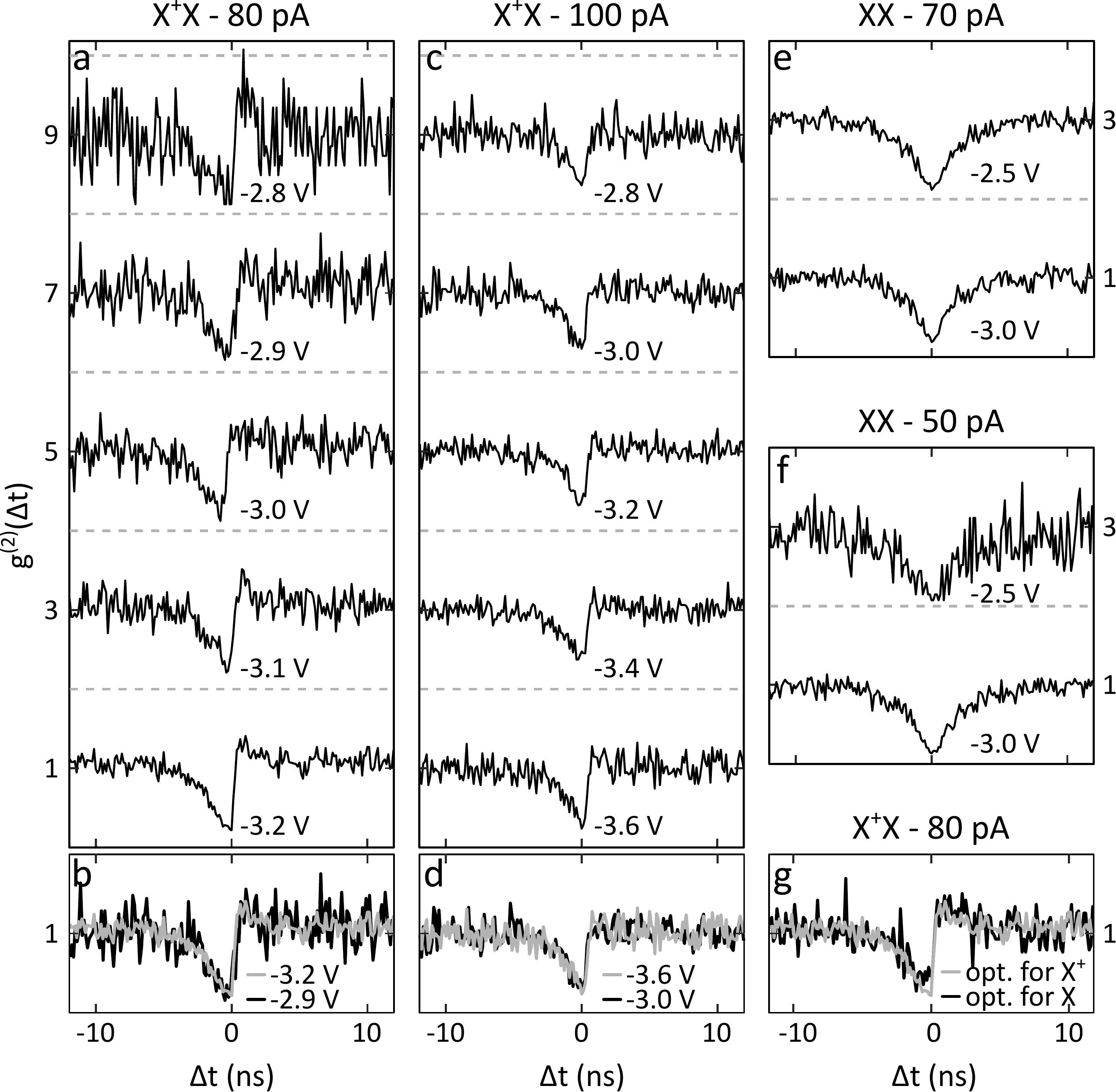}
    \caption{\textbf{Tip-position and voltage-dependence of the correlation measurements.} (a--d) X$^+$X correlation at different bias voltages, recorded at $80\,\text{pA}$ (a,b) and $100\,\text{pA}$ (c,d), respectively. In b and d, correlation curves recorded at two different bias voltages are overlaid. (e, f) XX correlation at different bias voltages, recorded at $70\,\text{pA}$ (e) and $50\,\text{pA}$ (f). (g) X$^+$X correlation at different tip-positions, recorded at positions of maximized intensity of X$^+$ (black) and X (grey) at $-3\,\text{V}$, $80\,\text{pA}$. In a, c, e and f, the correlation curves are shifted by 2 for clarity. (a-d, g) Detection on one side of the optical setup with dichroic mirror and $800\,\text{nm}$ and $655\,\text{nm}$ bandpass. (e, f) Detection on both sides with $700\,\text{nm}$ shortpass - $650\,\text{nm}$ longpass combination and $655\,\text{nm}$ bandpass.}
    \label{FigS3}
\end{figure}

\noindent Whether a given many-body transition is energetically possible depends on the applied bias voltage. \Figref{FigS3_1} shows two STML spectra recorded at the same lateral position atop a ZnPc molecule, but at different bias voltages. At -3\,V both excitonic and trionic emission can be observed, while at -2.5\,V only the excitonic emission is visible. To investigate the influence of the applied bias voltage on the exciton and trion dynamics, we recorded the X$^+$X and XX correlation at different applied bias voltages and currents (\Figref{FigS3}). X$^+$X correlation measurements recorded between $-2.8\,\text{V}$ and $-3.6\,\text{V}$ at $80\,\text{pA}$ and $100\,\text{pA}$ are shown in \Figref{FigS3}a-d. In \Figref{FigS3}b and d, two typical curves are overlaid, showing that the curve progression is similar in both cases. This is also visible in the XX correlation (\Figref{FigS3}e, f) recorded at $70\,\text{pA}$ and $50\,\text{pA}$. Note that, at $-2.5\,\text{V}$, the trion cannot be formed by consecutive charge-transfer from tip and substrate. \\

In addition, we recorded the X$^+$X correlation with the tip positioned at two different positions atop the molecule. To this end, we used the atom tracking feature to maximize X$^+$ emission and X emission, respectively. The corresponding correlation at the positions optimized for X$^+$ and X, shown in \Figref{FigS3}g, are not affected by the change in tip position.\\
\newpage

\newpage

\section*{S3 -- Theoretical model}
\noindent To interpret the correlation functions measured in the experiment, we design a theoretical model based on rate equations simulating the dynamics of the populations of relevant many-body states of the molecule. The dynamics is driven by a set of rates reflecting the charging and discharging processes mediated by the tip and the substrate, and the direct exciton (trion) decay rates mediated by their interaction with the picocavity plasmons. We include in the model the states of the neutral molecule, the molecular anion, and the molecular cation and dication. The states of the neutral molecule included in the model are the ground state $S_0$, the triplet excited state $T_1$ (six times degenerate), and the first singlet excited state $S_1$ (two times degenerate). For the cation we consider the doublet ground state $D_0^+$ (doubly degenerate), and the excited doublet $D_1^+$ (four times degenerate). The dication is represented by the singlet ground state $S_0^{2+}$ (non-degenerate) and the anion by its doublet ground state $D_0^{-}$ (four times degenerate). 
The state degeneracies are derived from (i) the spin degeneracy, and (ii) the orbital degeneracy of the lowest unoccupied molecular orbital (LUMO) of the neutral molecule. The latter is a result of the D$_{\rm 4h}$ symmetry of the molecule in the $S_0$ state.\\ 

More practically, we need to establish a mathematical framework that allows us to derive the selection rules governing the respective transition rates between the states. To that end, the many-body states of the molecule can be approximately defined with the help of fermionic electron creation and annihilation operators $c_i^\dagger$ ($c_i$) putting (removing) an electron into (from) a spin orbital $i$. We start by defining the cation ground state using the bra-ket notation $|S_0^{2+}\rangle\equiv|0\rangle$ as the least occupied state of the molecule. Assuming that the molecular orbitals can be approximately described using the orbitals of the neutral molecule regardless of the actual molecule's charge state, the remaining states can be approximately defined by adding electrons into the unoccupied orbitals of $|S_0^{2+}\rangle$. More explicitly, for the dication:
\begin{align}
    |S_0^{2+}\rangle &\equiv |0\rangle.\label{seq:state1}
\end{align}
For the cation we define the ground-state doublet
\begin{align}
    |D_0^{+}, \uparrow\rangle &\approx c^\dagger_{\rm H \uparrow}|0\rangle,\\
    |D_0^{+}, \downarrow\rangle &\approx c^\dagger_{\rm H \downarrow}|0\rangle,
\end{align}
and the pair of excited doublets
\begin{align}  
    |D_1^{+}, {\rm X}\uparrow\rangle &\approx c^\dagger_{\rm L_X \uparrow}|0\rangle,\\
    |D_1^{+}, {\rm X}\downarrow\rangle &\approx c^\dagger_{\rm L_X \downarrow}|0\rangle,\\
    |D_1^{+}, {\rm Y}\uparrow\rangle &\approx c^\dagger_{\rm L_Y \uparrow}|0\rangle\\
    |D_1^{+}, {\rm Y}\downarrow\rangle &\approx c^\dagger_{\rm L_Y \downarrow}|0\rangle.
\end{align}
For the neutral molecule we get the ground state
\begin{align}
    |S_0\rangle\approx c^\dagger_{\rm H\downarrow}c^\dagger_{\rm H\uparrow}|0\rangle,
\end{align}
the two sets of triplet states
\begin{align}
    |T_1, {\rm X}\uparrow\uparrow\rangle &\approx c^\dagger_{\rm L_X\uparrow}c^\dagger_{\rm H\uparrow}|0\rangle,\\
    |T_1, {\rm X}\downarrow\uparrow\rangle &\approx \frac{1}{\sqrt{2}} (c^\dagger_{\rm L_X\downarrow}c^\dagger_{\rm H\uparrow}+c^\dagger_{\rm L_X\uparrow}c^\dagger_{\rm H\downarrow})|0\rangle,\\
    |T_1, {\rm X}\downarrow\downarrow\rangle &\approx c^\dagger_{\rm L_X\downarrow}c^\dagger_{\rm H\downarrow}|0\rangle,
\end{align}
and 
\begin{align}
    |T_1, {\rm Y}\uparrow\uparrow\rangle &\approx c^\dagger_{\rm L_Y\uparrow}c^\dagger_{\rm H\uparrow}|0\rangle,\\
    |T_1, {\rm Y}\downarrow\uparrow\rangle &\approx \frac{1}{\sqrt{2}}(c^\dagger_{\rm L_Y\downarrow}c^\dagger_{\rm H\uparrow}+c^\dagger_{\rm L_Y\uparrow}c^\dagger_{\rm H\downarrow})|0\rangle,\\
    |T_1, {\rm Y}\downarrow\downarrow\rangle &\approx c^\dagger_{\rm L_Y\downarrow}c^\dagger_{\rm H\downarrow}|0\rangle,
\end{align}
and the two singlet excited states
\begin{align}
    |S_1, {\rm X}\rangle &\approx \frac{1}{\sqrt{2}}(c^\dagger_{\rm L_X\downarrow}c^\dagger_{\rm H\uparrow}-c^\dagger_{\rm L_X\uparrow}c^\dagger_{\rm H\downarrow})|0\rangle,\\
    |S_1, {\rm Y}\rangle &\approx \frac{1}{\sqrt{2}}(c^\dagger_{\rm L_Y\downarrow}c^\dagger_{\rm H\uparrow}-c^\dagger_{\rm L_Y\uparrow}c^\dagger_{\rm H\downarrow})|0\rangle.
\end{align}
Finally, the ground state of the anion is four times degenerate:
\begin{align}
    |D_0^-, {\rm X}\uparrow\rangle &\approx c^\dagger_{\rm L_X\uparrow}c^\dagger_{\rm H\downarrow}c^\dagger_{\rm H\uparrow}|0\rangle,\\
    |D_0^-, {\rm X}\downarrow\rangle &\approx c^\dagger_{\rm L_X\downarrow}c^\dagger_{\rm H\downarrow}c^\dagger_{\rm H\uparrow}|0\rangle,\\
    |D_0^-, {\rm Y}\uparrow\rangle &\approx c^\dagger_{\rm L_Y\uparrow}c^\dagger_{\rm H\downarrow}c^\dagger_{\rm H\uparrow}|0\rangle,\\
    |D_0^-, {\rm Y}\downarrow\rangle &\approx c^\dagger_{\rm L_Y\downarrow}c^\dagger_{\rm H\downarrow}c^\dagger_{\rm H\uparrow}|0\rangle.\label{seq:state7D}
\end{align}
Any electron-transfer event must be governed by an operator of the form $T=\sum_i t_i (c_{i\downarrow}+c_{i\uparrow})$ or $T^\dagger$. To derive the selection rules linked with charging and discharging events we consider a particularly simple form of this operator by setting all $t_i=1$ and evaluate the matrix elements of this operator between all pairs of the many-body states. \\
The selection rules governing the charging and discharging processes are summarized in matrices ${\bf M}_{AB}=|\langle A| T|B\rangle|^2 $ connecting pairs of many-body states $A, B\in\{ S_0^{2+}, D_0^+, D_1^+, S_0, T_1, S_1, D_0^- \}$ from the respective degenerate manifolds (in the order given in Eq.\,\eqref{seq:state1} to Eq.\,\eqref{seq:state7D}). The nonzero matrices are 
\begin{align}
    {\bf M}_{S_0^{2+}D_0^+}=
    \begin{bmatrix}
1 & 1 
\end{bmatrix}\label{seq:M1}
\end{align}
between the dication ground state and the ground cation state,
\begin{align}
    {\bf M}_{S_0^{2+}D_1^+}=
    \begin{bmatrix}
1 & 1 & 1 & 1 
\end{bmatrix}
\end{align}
between the dication ground state and the excited cation states,
\begin{align}
    {\bf M}_{D_0^{+}S_0}=
    \begin{bmatrix}
 1 \\
 1
\end{bmatrix}
\end{align}
between the cation ground state and the neutral ground state,
\begin{align}
    {\bf M}_{D_0^{+} T_1 }=
    \begin{bmatrix}
1 & \frac{1}{{2}} & 0 & 1 & \frac{1}{{2}} & 0 \\
0 & \frac{1}{{2}} & 1 & 0 & \frac{1}{{2}} & 1
\end{bmatrix}
\end{align}
between the cation ground states and the neutral triplet states,
\begin{align}
{\bf M}_{D_1^{+} T_1}=
\begin{bmatrix}
1 & \frac{1}{{2}} & 0 & 0 & 0 & 0 \\
0 & \frac{1}{{2}} & 1 & 0 & 0 & 0 \\
0 & 0 & 0 & 1 & \frac{1}{{2}} & 0 \\
0 & 0 & 0 & 0 & \frac{1}{{2}} & 1 
\end{bmatrix}
\end{align}
between the cation excited states and the neutral triplet states,
\begin{align}
{\bf M}_{D_0^{+} S_1}=
\begin{bmatrix}
\frac{1}{{2}} & \frac{1}{{2}} \\
\frac{1}{{2}} & \frac{1}{{2}} 
\end{bmatrix}
\end{align}
between the cation ground states and the neutral excited singlets,
\begin{align}
{\bf M}_{S_0 D_0^-}=
\begin{bmatrix}
1 &
1 &
1 &
1
\end{bmatrix}
\end{align}
between the neutral ground state and the anion ground state,
\begin{align}
{\bf M}_{T_1 D_0^-}=
\begin{bmatrix}
1 & 0 & 0 & 0\\
\frac{1}{{2}} & \frac{1}{{2}} & 0 & 0 \\
0 & 1 & 0 & 0\\
0 & 0 & 1 & 0\\
0 & 0 & \frac{1}{{2}} & \frac{1}{{2}}\\
0 & 0 & 0 & 1
\end{bmatrix}
\end{align}
between the neutral triplet states and the anion ground state, and finally
\begin{align}
{\bf M}_{S_1 D_0^-}=
\begin{bmatrix}
\frac{1}{{2}} & \frac{1}{{2}} & 0 & 0\\
 0 & 0 & \frac{1}{{2}} & \frac{1}{{2}} 
\end{bmatrix}\label{seq:M11}
\end{align}
between the neutral singlet excited states and the anion ground state. These matrices represent transitions where an electron is removed from the molecule. The matrices representing the reverse process can be obtained by transposing the matrices in Eq.\,\eqref{seq:M1} to Eq.\,\eqref{seq:M11}. \\

To derive the selection rules governing the respective charge-conserving transition rates [the exciton $S_1\to S_0$ ($X$), and the trion $D_1^+\to D_0^+$ ($X^+$) transitions], we consider that direct decay of the molecule occurs only between a pair of states of the same spin because it must be governed by a single-particle operator of the form $D=\sum_{ij}A_{ij}({c^\dagger_{i\downarrow} c_{j\downarrow}}+{c^\dagger_{i\uparrow} c_{j\uparrow}})$, where we explicitly defined the spin index. Again, we set $A_{ij}=1$ and obtain the following two matrices defined as ${\bf M}_{AB}=|\langle A| D |B \rangle|^2$, with $A, B\in \{ D_0^+, D_1^+, S_0, S_1 \}$.
\begin{align}
{\bf M}_{D_0^+D_1^+}=
\begin{bmatrix}
1 & 0 & 1 & 0 \\
0 & 1 & 0 & 1
\end{bmatrix}
\end{align}
and
\begin{align}
{\bf M}_{S_0S_1}=
\begin{bmatrix}
2 & 2 
\end{bmatrix}.
\end{align}
The matrices describing the excitation processes are in principle defined as the transpose of the matrices above, but we do not include the excitation processes in our model (see further discussion below). \\

We next build the rate equations for the populations $N_i$ of the many-body state $i$ that can be jointly cast into the vector $\textbf{N}$ containing the populations of all 20 states defined above [maintaining again the order given in Eq.\,\eqref{seq:state1} to Eq.\,\eqref{seq:state7D}]. The rate equation then acquires the form:
\begin{align}
    \dot{\textbf{N}}= {\mathcal{M}}\textbf{N},\label{seq:req}
\end{align}
where 
\begin{align}
&{\mathcal{M}}=\nonumber\\
&\begin{bmatrix}
-{\bf \Gamma}_{S_0^{2+}S_0^{2+}} & {\bf \Gamma}_{S_0^{2+}D_0^+} & {\bf \Gamma}_{S_0^{2+}D_1^+}& \textbf{0} & \textbf{0} & \textbf{0} & \textbf{0} \\
{\bf \Gamma}_{D_0^+S_0^{2+}} & -{\bf \Gamma}_{D_0^{+}D_0^{+}} & {\bf \Gamma}_{D_0^+D_1^{+}} & {\bf \Gamma}_{D_0^+S_0} & {\bf \Gamma}_{D_0^+T_1} & {\bf \Gamma}_{D_0^+ S_1} & \textbf{0} \\
{\bf \Gamma}_{D_1^+ S_0^{2+}} & \textbf{0} & -{\bf \Gamma}_{D_1^{+}D_1^{+}} & {\bf \Gamma}_{D_1^+ S_0} & {\bf \Gamma}_{D_1^+ T_1} & {\bf \Gamma}_{D_1^+ S_1} & \textbf{0} \\
\textbf{0} & {\bf \Gamma}_{S_0D_0^{+}} & {\bf \Gamma}_{S_0D_1^{+}} & -{\bf \Gamma}_{S_0S_0} & \textbf{0} & {\bf \Gamma}_{S_0S_1} & {\bf \Gamma}_{S_0D_0^-} \\
\textbf{0} & {\bf \Gamma}_{T_1D_0^{+}} & {\bf \Gamma}_{T_1D_1^{+}} &  \textbf{0} & -{\bf \Gamma}_{T_1T_1}  & \textbf{0} & {\bf \Gamma}_{T_1D_0^-} \\
\textbf{0} & {\bf \Gamma}_{S_1D_0^{+}} & {\bf \Gamma}_{S_1D_1^{+}} &  \textbf{0} & \textbf{0}  & -{\bf \Gamma}_{S_1S_1} & {\bf \Gamma}_{S_1D_0^-} \\
\textbf{0} & \textbf{0} & \textbf{0} &  {\bf \Gamma}_{D_0^-S_0} & {\bf \Gamma}_{D_0^-T_1}  & {\bf \Gamma}_{D_0^-S_1} & -{\bf \Gamma}_{D_0^-D_0^-} 
\end{bmatrix},
\end{align}
and ${\bf \Gamma}_{ij}=\gamma_{ij}{\bf M}_{ij}$ for $i\neq j$, and ${\bf \Gamma}_{ii}=\gamma_{ii}\textbf{1}$ with $\textbf{1}$ being an identity matrix of dimensions corresponding to the degeneracy of the state $i$, and $\textbf{0}$ are zero matrices of corresponding dimensions. We note that we only included charge- and spin-conserving deexcitation processes but neglected the inverse counterparts. We therefore assume that the molecule is driven to its optically excited states purely by charging and discharging via the sequential tunneling mechanism. The diagonal rates are calculated as a sum of all decay channels leading from state $i$. As a result, the sum of each column of the matrix $\mathcal{M}$ is zero. The rates $\gamma_{ij}$ are treated as parameters as we detail below.\\

Charge and spin-conserving decay rates (photon emission events) are for simplicity described by two parameters whose value is fixed and independent of the tip-sample distance. Namely we choose $\hbar\gamma_{S_0S_1}=0.5$\,meV and $\hbar\gamma_{D_0^+D_1^+}=1$\,meV. This is an approximation which does not strongly influence the overall charge dynamics (steady state, nor photon correlation functions) derived from the rate equations, provided that these rates are chosen to be much larger than the charging and discharging rates. Indeed, this assumption is justified since the charge and spin-conserving decay rates are enhanced by the presence of the picocavity plasmons and reach units of meV \cite{Roslawska2022PRX,Dolezal2024NanoLett}. On the other hand, for the tunneling configuration in question, the charging rates are usually several orders of magnitude smaller ($\sim 0.1 \,\mu{\rm eV}$) as they correspond to slower processes.  \\

The rates governing the charging and discharging processes are parametrized depending on whether they are mediated by the substrate or the tip. The substrate-mediated processes are treated as independent of the tip-sample distance, and their values are defined in Table\,\ref{stab:rates}. On the other hand, the tip-mediated processes are treated as dependent on the tip-sample distance and are described by the following formula:
\begin{align}
    \gamma_{ij}=\gamma^0_{ij}\exp(-2\kappa_{ij}L_{\rm eff}),\label{seq:tiprates}
\end{align}
where $\gamma^0_{ij}$ is a constant (defined in Table\,\ref{stab:rates}), $\kappa_{ij}=\sqrt{2m_e|E_i^{E_{\rm vac}}-E_j^{E_{\rm vac}}|}$ is a decay rate with $m_e$ being the electron mass, and $E_i^{E_{\rm vac}}$ being the energy of the respective many-body state assuming that during ionization (electron capture) the emitted (captured) electron is inserted into (extracted from) the vacuum level at rest. These energies are summarized in Table\,\ref{stab:energies} and compared to the energies $E_i^{\mu}$ shown in the diagram in Fig. 2 of the main paper. More practically, the two sets of energies can be transformed into each other by using the formula $E_i^\mu=E_i^{E_{\rm vac}}-WQ_i$, where $W$=3.58 eV is the substrate work function \cite{Imada2018prb} and $Q_i$ is the total charge of the state $i$ measured in units of the electron charge. The effective distance $L_{\rm eff}$ is a fitting parameter that is physically motivated by the distance between the molecule and the tip. Last, we note that the constant parameters are chosen as an educated guess based on available experimental data in this paper and in the literature \cite{Steurer2014ApplPhysLett,Kaiser2023NatCommun}. Furthermore, we do not attempt to precisely fit the experimental data by varying all the rates present in the model. Instead, for simplicity, we choose an identical value for all the substrate-mediated rates, and only two distinct values for the $\gamma^0_{ij}$ parameters (distinguishing which orbital is involved in the tunneling process).  \\
We also note that the rate $\gamma_{S_0 D_0^-}=\gamma^{\rm T}_{S_0 D_0^-}+\gamma^{\rm S}_{S_0 D_0^-}$ has two components, $\gamma^{\rm S}_{S_0 D_0^-}$ mediated by the substrate and treated as a constant, and $\gamma^{\rm T}_{S_0 D_0^-}$ mediated by the tip and thus following Eq.\,\eqref{seq:tiprates}. 

\begin{table}
\centering
\caption{Values of parameters used in the rate-equation model. Notice that the tip-mediated rates are calculated using Eq.\,\eqref{seq:tiprates}, while the substrate rates are directly given by the values listed here.}
\vspace{0.3cm}
\begin{tabular}{c|c||c|c}
\multicolumn{2}{c||}{Substrate-mediated} & \multicolumn{2}{c}{Tip-mediated} \\
\hline
Parameter & Value (eV) & Parameter & Value (eV) \\
\hline
$\hbar\gamma_{D_0^+S_0^{2+}}$ & 9.924$\times10^{-7}$ & $\hbar\gamma^0_{S_0^{2+}D_0^+}$ & 1 \\
$\hbar\gamma_{D_1^+S_0^{2+}}$ & 9.924$\times10^{-7}$ & $\hbar\gamma^0_{D_0^+S_0}$ & 1 \\
$\hbar\gamma_{S_0 D_0^{+}}$  & 9.924$\times10^{-7}$ & $\hbar\gamma^0_{D_0^+T_1}$ & 5 \\
$\hbar\gamma_{T_1 D_0^{+}}$  & 9.924$\times10^{-7}$ & $\hbar\gamma^0_{D_1^+T_1}$ & 1 \\
$\hbar\gamma_{S_1 D_0^{+}}$  & 9.924$\times10^{-7}$ & $\hbar\gamma^0_{S_0D_0^-}$ & 5 \\
$\hbar\gamma_{D_0^- T_1}$    & 9.924$\times10^{-7}$ & $\hbar\gamma^0_{T_1D_0^-}$ & 1 \\
$\hbar\gamma_{ S_0D_0^-}^{\rm S}$ & 9.924$\times10^{-7}$ & $\hbar\gamma^0_{S_1D_0^-}$ & 1  
\end{tabular}\label{stab:rates}
\end{table}

\begin{table}
\centering
\caption{Energies of many-body states used in the model. We set the energy of the ground state of the neutral molecule $E_{S_0}=E_{S_0}^{\mu}=E_{S_0}^{E_{\rm vac}}=0$\,eV for convenience and without loss of generality. The energies are listed as $E_i^{\mu}$ and $E_i^{E_{\rm vac}}$, assuming that electrons are exchanged between the molecule and the substrate, or between the molecule and the environment, \textit{i.e.} vacuum. More explicitly, we assume that when an electron is exchanged between the molecule and the substrate, in the substrate it has the energy of the substrate electrochemical potential $\mu$, and that when an electron is exchanged between the molecule and the environment, in the environment it has the vacuum energy $E_{\rm vac}$ at rest. In the latter case, the energy differences have the meaning of ionization energies and electron affinities.}
\vspace{0.3cm}
\begin{tabular}{c||c|c}
State & $E_i^{\mu}$ (eV) & $E_i^{E_{\rm vac}}$ (eV) \\
\hline
$E_{S_0}$ & 0 & 0 \\
$E_{T_1}$ & 1.20 & 1.20 \\
$E_{S_1}$ & 1.89 & 1.89 \\
$E_{D_0^+}$ & 2.10 & 5.68 \\
$E_{D_1^+}$ & 3.62 & 7.2 \\
$E_{S_0^{2+}}$ & 4.60 & 11.76 \\
$E_{D_0^{-}}$ & 0.80 &  -2.78
\end{tabular}\label{stab:energies}
\end{table}

Having defined the rate equation [Eq.\,\eqref{seq:req}], we solve it to obtain the steady state of the system by setting the time derivative equal to zero. The steady-state solution is then the zero-eigenvalue eigenvector normalized such that the populations sum up to unity. The steady-state solution can be used to evaluate the tunneling current $I$ as a sum of charging rates mediated by the tip that are weighted by the steady-state populations of the corresponding initial many-body states. More explicitly:
\begin{align}
    \frac{\hbar}{e^2}I&=2\gamma_{D_0^+S_0}N_{S_0}+(\gamma_{D_0^+T_1}+\gamma_{D_1^+T_1})\sum_{i} N_{T_1,i}\nonumber\\
    &+(0.5\gamma_{S_1D_0^-}+1.5\gamma_{T_1D_0^-}+\gamma^{\rm T}_{S_0D_0^-})\sum_{i}N_{D_0^-,i} + \gamma_{S_0^{2+}D_0^+}\sum_i N_{D_0^+,i},
\end{align}
where the summation runs over the degenerate states. \\

Using the rate equations and choosing the right initial conditions we can also evaluate the photon-photon correlation functions. In particular, the following correlation functions are required to be evaluated:
\begin{align}
    g^{(2)}_{\rm XX}(\tau)&=\frac{\langle\sigma_{X}^\dagger(t)\sigma_{X}^\dagger(t+\tau)\sigma_{X}(t+\tau)\sigma_{X}(t)\rangle}{N_{S_1,{\rm X}}^2},\\
    g^{(2)}_{\rm X^+X^+}(\tau)&=\frac{\langle\sigma_{X^+}^\dagger(t)\sigma_{X^+}^\dagger(t+\tau)\sigma_{X^+}(t+\tau)\sigma_{X^+}(t)\rangle}{N_{D_1^+,{\rm X}\uparrow}^2},\\
    g^{(2)}_{\rm XX^+}(\tau)&=\frac{\langle\sigma_{X}^\dagger(t)\sigma_{X^+}^\dagger(t+\tau)\sigma_{X^+}(t+\tau)\sigma_{X}(t)\rangle}{N_{S_1,{\rm X}} N_{D_1^+,{\rm X}\uparrow}},\\
    g^{(2)}_{\rm X^+X}(\tau)&=\frac{\langle\sigma_{X^+}^\dagger(t)\sigma_{X}^\dagger(t+\tau)\sigma_{X}(t+\tau)\sigma_{X^+}(t)\rangle}{N_{S_1,{\rm X}} N_{D_1^+,{\rm X}\uparrow}},
\end{align}
with $g^{(2)}_{\rm XX^+}(-\tau)=g^{(2)}_{\rm X^+X}(\tau)$ and
\begin{align}
    \sigma_X&=| S_0 \rangle \langle S_{1}, {\rm X} |,\\
    \sigma_{X^+}&=| D_0^+, \uparrow \rangle \langle D_1^+, {\rm X} \uparrow |,
\end{align}
where for specificity we picked the excitations polarized along $x$ and for the doublet we consider only the spin up state. Since our model treats all the degenerate states equivalently, this choice is general. \\

The quantum regression theorem states that two-time correlation functions of the type $\langle A(0)B(\tau)A^\dagger(0)\rangle$ obey the dynamics of the mean value of the operator $B(\tau)$, but impose an initial condition to the dynamics given by the steady-state value $\langle ABA^\dagger \rangle$. In the case of our correlation functions, we therefore solve the dynamics of the corresponding populations $N_i(\tau)\equiv\langle \sigma_{ii}(\tau) \rangle$ ($\sigma_{ii}=|i\rangle\langle i|$) defined by the rate equation Eq.\,\eqref{seq:req} with the initial values defined by $\langle A \sigma_{ii} A^\dagger\rangle$, where $A$ is one of the operators $\sigma_{X}^\dagger$ and $\sigma_{X^+}^\dagger$. Practically, this means that we solve the dynamics of the populations assuming that the system is initially fully in the $|S_0\rangle$ state for $g^{(2)}_{\rm XX}(\tau)$ and $g^{(2)}_{\rm XX^+}(\tau)$ (in the $|D_0^+, \uparrow\rangle$ state for $g^{(2)}_{\rm X^+X^+}(\tau)$ and $g^{(2)}_{\rm X^+X}(\tau)$) and to obtain the correlation function we evaluate the populations of the states $|{S}_1, {\rm X}\rangle$ and $|{D}_1^+, {\rm X}\uparrow\rangle$ ($|{D}_1^+, {\rm X}\uparrow\rangle$ and $|{S}_1, {\rm X}\rangle$), respectively. Finally, to account for the effect of limited instrumental time resolution we convolve the calculated correlation functions with a Gaussian of width $\sigma=250$ ps.

\newpage

\section*{S4 -- Current-dependent transition dynamics}

\begin{figure}[!ht]
    \centering
    \includegraphics{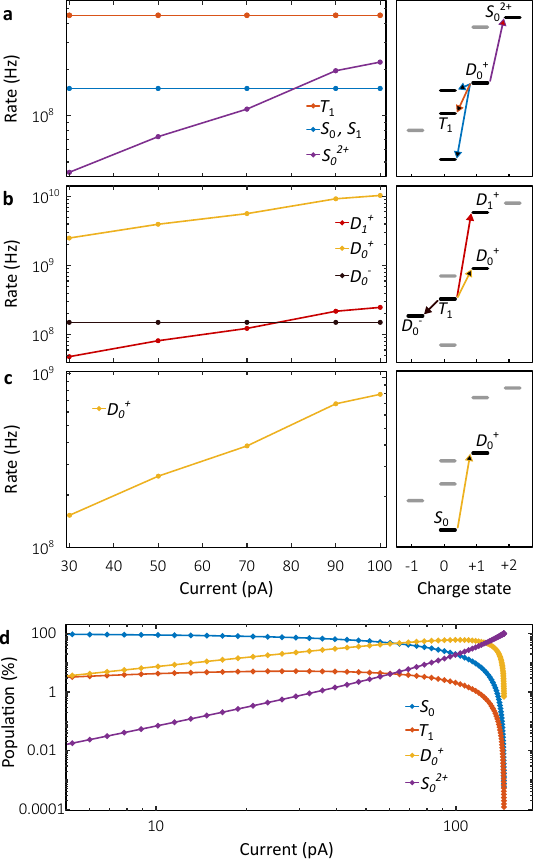}
    \caption{\textbf{Current-dependent charge-state transition dynamics.} (a--c) Calculated current-dependent transition rates from $D_0^+$ (a) and $T_1$ (b) and $S_0$ (c) as initial state, respectively. The solid lines are guides to the eye. The panels on the right show the corresponding possible transitions in a many-body scheme. (d) Calculated relative population of $S_0$, $T_1$, $D_0^+$ and $S_0^{2+}$ as a function of tunnel current.}
    \label{FigS4}
\end{figure}

\addcontentsline{toc}{subsection}{Current-dependent transition rates}
\noindent\textbf{Current-dependent transition rates}\\
\begin{table}[!b]
\centering
\caption{Range of lifetimes of the mediating states as extracted from the model's fit to the data. The left and right lifetimes correspond to the lifetimes at 30 pA and 100 pA, respectively.}
\vspace{0.3cm}
\begin{tabular}{c||c}
State & Lifetime (ns)\\
\hline
$S_0$ & 6.53 -- 1.32 \\
$T_1$ & 0.37 -- 0.09 \\
$S_1$ & $<$ 0.02 \\
$D_0^+$ & 1.26 -- 1.02 \\
$D_1^+$ & $<$ 0.01 \\
$S_0^{2+}$ & 1.11 -- 1.11 \\
$D_0^{-}$ & 0.02 -- 0.01
\end{tabular}\label{stab:lifetimes}

\centering
\caption{Range of timescales at which the initial states $S_0$, $T_1$ and $D_0^+$ are depopulated, as extracted from the model's fit to the data. The left and right timescales correspond to the average time at which the corresponding transitions occur at 30 pA and 100 pA, respectively.}
\vspace{0.3cm}
\begin{tabular}{c||c}
State & Timescale (ns)\\
\hline
$S_0\rightarrow D_0^+$ & 6.53 -- 1.32\\
\hline
$T_1\rightarrow D_0^+$ & 0.40 -- 0.10\\
$T_1\rightarrow D_1^+$ & 20.83 -- 4.03\\
$T_1\rightarrow D_0^-$ & 6.63 -- 6.63\\
\hline
$D_0^+\rightarrow S_0^{2+}$ & 23.37 -- 4.47\\
$D_0^+\rightarrow S_1$ & 6.63 -- 6.63\\
$D_0^+\rightarrow T_1$ & 2.21 -- 2.21\\
$D_0^+\rightarrow S_0$ & 6.63 -- 6.63
\end{tabular}\label{stab:rates_list}
\end{table}
From the model, we can assess the transition rates between the individual states that are relevant for the excited state formation dynamics. \Figref{FigS4}a--c shows the current-dependence of all non-zero transition rates from the initial states $D_0^+$, $T_1$ and $S_0$, respectively. The lifetimes of the different many-body states (Table\,\ref{stab:lifetimes}) are composed of the rates at which the corresponding states are depopulated (Table\,\ref{stab:rates_list}). Thus, the fastest transition rate from a given state will dominate its lifetime. For $D_0^+$ (\Figref{FigS4}a), this is the sample-mediated transition to $T_1$ within the probed current range. Consequently, the lifetime of $D_0^+$ stays practically constant within that range. The transition dominating the $T_1$ lifetime, on the other hand, is the tip-mediated transition to $D_0^+$ (\Figref{FigS4}b), such that the $T_1$ lifetime decreases appreciably as a function of current. $S_0$ can only be depopulated by a tip-mediated transition to $D_0^+$ (\Figref{FigS4}c), and thus also the $S_0$ lifetime changes with current. Although $D_0^+$ is populated in a tip-mediated process from both $S_0$ and $T_1$, the $S_0\rightarrow D_0^+$ transition is significantly slower than the $T_1\rightarrow D_0^+$ transition. This relates to an appreciably smaller tunneling barrier height for the latter process. \\
\vspace{1cm}

\addcontentsline{toc}{subsection}{Current-dependent populations}
\noindent\textbf{Current-dependent populations}\\
We extract the average population of the mediating states at different tunnel currents by extrapolating the transition rates from the fits to the correlation curves to smaller or larger tip-molecule distances. \Figref{FigS4}d shows the calculated populations of $S_0$, $T_1$, $D_0^+$ and $S_0^{2+}$ as a function of current. For currents $< 60\;\text{pA}$, the molecule is neutral most of the time. With increasing current the time during which it is charged increases, and above 125 pA the molecule is  mostly doubly charged.

\bibliographystyle{naturemag}
\bibliography{HBT_ref}